\documentclass[journal]{IEEEtran}
\usepackage[utf8]{inputenc}
\usepackage{amsmath}
\usepackage{amssymb}
\usepackage{amsfonts}
\usepackage{mathtools}
\usepackage{bm}
\usepackage{graphicx}
\usepackage{booktabs}
\usepackage{multirow}
\usepackage{tabularx}
\usepackage{array}
\usepackage{tabularx}
\usepackage{algorithm}
\usepackage{algpseudocode}
\usepackage{url}
\usepackage{tcolorbox}
\usepackage{fvextra}
\usepackage{xcolor}
\usepackage{cite}
\usepackage{balance}
\usepackage[hidelinks]{hyperref}
\hypersetup{
  colorlinks=true,
  linkcolor=blue,
  citecolor=blue,
  urlcolor=blue,
  filecolor=blue
}
\usepackage[normalem]{ulem}

\begin{document}
\title{Swarm-Driven Multi-Agent Reasoning for Smart City Security}

\author{
Saeid~Jamshidi,
Carol~Fung,
Kawser~Wazed~Nafi,
Foutse~Khomh
\thanks{
K. W. Nafi and F. Khomh is with the SWAT Laboratory,
Polytechnique Montréal, Montréal, QC, Canada
(e-mail: \{kawser.wazed-nafi, foutse.khomh\}@polymtl.ca).
}
\thanks{
S. Jamshidi  and C. Fung are with the Concordia Institute for Information Systems Engineering (CIISE),
Concordia University, Montréal, QC, Canada
(e-mail: \{saeid.jamshidi, carol.fung\}@concordia.ca).
}
}

\maketitle

\begin{abstract}
Modern smart cities operate as complex, interconnected cyber-physical ecosystems in which thousands of heterogeneous devices continuously exchange data and control commands. In these environments, advanced threats often differ from conventional isolated incidents; i.e., a low-rate scan of traffic sensors, irregular credential usage across edge devices, protocol misuse, and delayed lateral movement across gateways may each remain below local alert thresholds but collectively indicate a coordinated multi-stage campaign. Therefore, security in smart cities remains not only an attack detection problem but also a reasoning problem under uncertainty, partial observability, and adversarial manipulation. In this work, we present \textit{TPSC-Sec}, an LLM-based multi-agent approach for stable and reliable security reasoning in smart cities. In contrast to a single-agent model, which may overlook distributed indicators and produce unstable interpretations, TPSC-Sec decomposes security analysis across specialized agents that examine traffic behavior, protocol interactions, identity usage, and temporal attack progression. These agents generate independent threat hypotheses from partial observations, which are then aggregated through the proposed \textit{Threat-Pheromone Swarm Consensus (TPSC)} mechanism. TPSC tracks hypothesis-support dynamics through reinforcement, contradiction handling, and temporal consistency, enabling competing threat interpretations to converge toward a stable collective decision. We further introduce \textit{Adaptive Verified TPSC (AV-TPSC)}, which adds verification-aware calibration, context-sensitive weighting, and disagreement-adaptive control to reduce unsupported LLM outputs and reasoning inconsistencies under adversarial conditions. Experimental results over 500 runs show that \textit{TPSC-Sec} achieves stable consensus formation with a high acceptance rate of $0.97 \pm 0.02$, strong hypothesis-support concentration ($>0.99$), and a consensus margin of $2.08 \pm 0.21$. The system maintains low aggregate risk ($0.23 \pm 0.04$), high inter-agent agreement ($0.82 \pm 0.06$), and strong support-quality correlation ($r=0.93$). Adaptive agent selection further reduces the number of active agents by 50\% while improving overall system fitness by 11.6\%. These results demonstrate that \textit{TPSC-Sec} enables robust, interpretable, and computationally efficient security reasoning for adversary-resilient smart-city environments.
\end{abstract}


\maketitle

\section{Introduction}
\label{Introduction}
The rapid evolution of smart city infrastructure has enabled the large-scale deployment of Internet of Things (IoT) systems that integrate sensing, communication, and control across critical services such as transportation, energy, healthcare, and public safety \cite{firoozi2025impact,khan2026xai,ansari2025iot}. These systems form highly interconnected cyber-physical ecosystems composed of heterogeneous devices, edge gateways, cloud services, and software controllers operating over dynamic and resource-constrained networks \cite{barbone2025device,mishra2025cyber,lin2025fault}. While this integration enables intelligent and autonomous city-scale operations, it also expands the attack surface and introduces security challenges that are difficult to address using isolated detection mechanisms \cite{tanimu2025addressing,zhukabayeva2025cybersecurity}.
In smart cities, security incidents often emerge from temporally distributed and cross-domain behaviors rather than from a single, obvious malicious event. Devices communicate using diverse protocols such as MQTT, CoAP, HTTP, Modbus/TCP, Zigbee, Bluetooth Low Energy, and cellular IoT protocols, generating heterogeneous telemetry, protocol traces, authentication records, and system logs across traffic, identity, protocol, and temporal dimensions. A conventional isolated attack may appear as a direct denial-of-service (DoS) event against a gateway, whereas an advanced multi-stage attack may begin with low-rate scanning of traffic sensors, continue with subtle identity anomalies such as credential reuse from atypical devices and geographically inconsistent access patterns, exploit a protocol misuse pattern, and later perform lateral movement across service gateways. Each step may remain below local detection thresholds, but the combined sequence can indicate a coordinated multi-stage campaign. Therefore, security in smart cities must be treated not only as an attack detection task but also as a reasoning task under uncertainty, partial observability, and adversarial manipulation \cite{hossain2025intergraph,zografopoulos2025cyber,shaukat2025review,alotaibi2025review,ponnumani2026multi}.
Existing IoT security approaches have made important progress in optimization, anomaly detection, and intrusion classification \cite{shaukat2025review,alotaibi2025review,wu2024deep,yang2024continual}. Learning-based methods, including machine learning (ML) and deep learning (DL) models, can achieve strong detection accuracy on benchmark datasets, while swarm intelligence and multi-agent reinforcement learning provide scalable mechanisms for decentralized optimization, scheduling, task allocation, and resource-aware decision-making in dynamic IoT environments \cite{zhou2022swarm,anuraj2024dynamic,dimos2025survey,taher2022novel,lu2025bpso}. Swarm intelligence is particularly attractive because it supports distributed coordination without relying on a single centralized controller. However, most existing swarm-based IoT security methods apply swarm principles mainly to numerical optimization, feature selection, routing, and resource allocation. They do not explicitly reason over security-relevant telemetry and logs, resolve conflicting threat interpretations, and model how adversarial actions influence the reasoning process itself. Similarly, deep learning-based intrusion detection system (IDS) methods often rely on statistical feature representations and provide limited interpretability when attacks are fragmented across multiple devices, protocols, and time windows \cite{wu2024deep,yang2024continual}.\\
Recent advances in Large Language Models (LLMs) and multi-agent systems introduce new opportunities for semantic security reasoning and collaborative inference \cite{feng2025heterogeneous,zhu2025swarm,hashem2024distributed}. LLM-based agents can interpret heterogeneous evidence, explain protocol misuse, connect weak indicators across time, and generate structured hypotheses about multi-stage attacks. However, deploying LLM agents in security-critical IoT environments remains challenging because individual agents may hallucinate, overfit to incomplete evidence, produce inconsistent interpretations, and be manipulated by adversarial inputs such as prompt injection, misleading logs, and corrupted telemetry \cite{kong2025survey,rezaei2025intelligent}. A single LLM agent also creates a reasoning bottleneck: it must process all evidence at once, may overlook weak distributed indicators, and lacks an internal mechanism for disagreement resolution. These limitations motivate a multi-agent design in which specialized agents analyze complementary evidence views, and a principled consensus mechanism stabilizes their outputs.
In addition, a key unresolved problem is the lack of a unified approach that integrates semantic reasoning, swarm-based consensus, and adversary-aware robustness for smart-city security. Existing approaches typically address only one part of this problem: deep learning models improve detection but lack transparent reasoning; classical swarm methods support distributed coordination but do not perform semantic threat interpretation; and multi-agent LLM systems support collaborative reasoning but often lack stable, verification-aware consensus. More importantly, current solutions rarely capture the dual impact of adversarial actions. In this paper, dual impact refers to the ability of an adversary to affect both: 1) the observable smart-city environment, by modifying traffic, protocol behavior, identity usage, and service availability; and 2) the internal reasoning process, by introducing misleading, incomplete, and contradictory evidence that fragments agent beliefs and weakens consensus. Modeling both effects is essential for reliable security reasoning in complex smart-city environments \cite{guo2023deep,javed2024robustness,xu2025fortifying,ogenyi2025comprehensive}.\\
To address this gap, we propose \textit{TPSC-Sec (Threat-Pheromone Swarm Consensus Security)}, an LLM-based multi-agent approach for adversary-resilient security reasoning in smart cities. In contrast to conventional detection-based approaches, \textit{TPSC-Sec} formulates security as a dynamic reasoning and belief aggregation problem rather than a static classification task. This formulation enables reasoning over incomplete, conflicting, and temporally distributed evidence, which is essential for identifying multi-stage and context-dependent attacks. Instead of relying on a monolithic LLM, \textit{TPSC-Sec} decomposes security analysis into specialized reasoning agents that examine traffic behavior, protocol interactions, identity consistency, and temporal attack progression. This specialization allows each agent to focus on a coherent evidence view while preserving diversity across the reasoning process. The outputs of these agents are integrated through the proposed \textit{Threat-Pheromone Swarm Consensus (TPSC)} mechanism. Inspired by pheromone-based collective behavior in swarm intelligence, particularly Ant Colony Optimization (ACO) \cite{lopez2025ant}, TPSC adapts swarm principles from numerical optimization to semantic belief aggregation. Hypotheses supported by consistent evidence are reinforced, while weak and contradictory hypotheses are attenuated through contradiction-aware inhibition and temporal evaporation. This allows the system to converge toward stable decisions even when evidence is incomplete and distributed across multiple observation windows. Ablation results indicate that the basic TPSC mechanism improves consensus stability compared with majority voting, but its performance degrades when agent outputs contain low-verification hypotheses, overconfident unsupported reasoning, and high inter-agent disagreement under adversarial manipulation. Therefore, we further introduce \textit{Adaptive Verified TPSC (AV-TPSC)}, which incorporates verification-aware calibration, context-sensitive weighting, and disagreement-adaptive control to reduce the propagation of unsupported LLM outputs and reasoning inconsistencies.
 The main contributions of this paper are summarized as follows:
\begin{itemize}
\item \textbf{Swarm-Guided Multi-Agent Security Reasoning:}
We introduce \textit{TPSC-Sec}, an LLM-based multi-agent approach that decomposes smart-city security analysis into specialized reasoning agents for traffic behavior, protocol interaction, identity consistency, and temporal attack progression.

\item \textbf{Pheromone-Based Consensus for Stable Threat Interpretation:}
We propose the \textit{Threat-Pheromone Swarm Consensus (TPSC)} mechanism, which adapts swarm intelligence to semantic belief aggregation by modeling reinforcement, contradiction, and temporal consistency among competing threat hypotheses.

\item \textbf{Verification-Aware Robustness Against Adversarial Reasoning Disruption:}
We develop \textit{Adaptive Verified TPSC (AV-TPSC)} to mitigate hallucinations, misleading evidence, and inter-agent inconsistency through verification-aware confidence calibration, context-sensitive weighting, and disagreement-adaptive contradiction control.

\item \textbf{Adversary-Aware Joint Modeling of System and Reasoning Dynamics:}
We formulate the dual impact of adversarial behavior on both observable smart-city operations and internal multi-agent belief formation, enabling reliable decision-making under partial observability and multi-stage attack conditions.
\end{itemize}

The remainder of this paper is organized as follows. Section~\ref{Related Work} reviews related work. 
Section~\ref{Threat Model} presents the threat model. Section~\ref{Methodology} describes the proposed solution. Section~\ref{Experimental Setup} outlines the experimental setup. Section~\ref{Experimental Results} reports the results. Section~\ref{Discussion} discusses the findings. Section~\ref{Limitations and Future Work} discusses limitations and future work. Section~\ref{Conclusion} concludes the paper.

\section{Related Work}
\label{Related Work}
This section reviews the foundations of \textit{TPSC-Sec} and the gap it addresses in smart city security.

\subsection{Swarm Intelligence for IoT Security and Resource-Aware Decision Making}
Swarm intelligence has been widely adopted in IoT networks to support distributed decision-making, scheduling efficiency, and security-aware optimization. In resource-constrained environments, swarm-enabled multi-agent reinforcement learning has been used to optimize task offloading under dynamic constraints on computation, budget, and energy \cite{zhou2022swarm}. Similarly, swarm-based scheduling methods have improved IoT device security under energy, funding, and timing constraints \cite{nizamudeen2023intelligent}. These studies demonstrate the suitability of swarm intelligence for decentralized coordination in large-scale IoT networks. However, existing swarm-based methods remain primarily allocation- and optimization-driven. They focus on efficiency improvement rather than semantic interpretation of adversarial behavior, resolution of conflicting evidence, and stabilization of reasoning under uncertainty.

\subsection{Swarm-Enhanced DL for Intrusion Detection}
Another line of research integrates swarm intelligence with ML and DL for IDS. Swarm-based DL classifiers improve feature selection and attack detection in multi-cloud IoT networks \cite{taher2022novel}. Hybrid approaches combining tunicate swarm optimization with recurrent neural architectures achieve high precision and accuracy on benchmark datasets \cite{lu2025bpso}. More recent work applies binary particle swarm optimization to automate neural architecture and hyperparameter design, yielding results on the Bot-IoT, ToN-IoT, Gas Pipeline, and SWaT datasets \cite{chen2026swarm}. These approaches strengthen numerical learning pipelines, but their decision processes remain statistical and representation-driven. They do not explicitly reason about semantic attack context, subsystem interactions, temporally distributed evidence, and inter-agent disagreement. As a result, they improve predictive performance but do not provide collective reasoning and adversary-aware consensus.

\subsection{Multi-LLM Collaboration and Swarm-Based Reasoning}
Recent research has explored collaboration among multiple LLMs and the use of swarm principles for reasoning coordination. Heterogeneous swarm approaches model multi-LLM systems as directed acyclic graphs and optimize agent roles and weights using particle swarm optimization \cite{feng2025heterogeneous}. In parallel, swarm-enhanced reasoning formulates reasoning as solution-space exploration, improving diversity and solution quality through density estimation and non-dominated sorting \cite{zhu2025swarm}. These studies demonstrate the feasibility of swarm-guided coordination among reasoning agents. However, their focus is mainly on performance optimization for general reasoning tasks. They do not address adversarial conflict, evolving threat beliefs, and uncertainty-aware security consensus in distributed IoT and smart-city environments.

\subsection{Fault Tolerance, Validation, and Reliability in LLM-Based IoT Security}
A related research direction focuses on improving the reliability of LLM-based systems through redundancy, critique, validation, auditing, and reliability-weighted consensus. FAIR-Swarm introduces these mechanisms to enhance robustness in scientific hypothesis generation \cite{roy2026fair}. This work highlights important challenges for multi-agent LLM systems, including hallucinations, failure propagation, and the need for structured validation. However, its design targets scientific discovery rather than IoT security intelligence, and it does not model dynamic threat propagation, contradiction-sensitive reinforcement, and temporally evolving adversarial evidence. More broadly, recent analyses show that LLM-based swarms can introduce significant computational overhead compared with classical swarm systems \cite{rahman2025llm}, limiting their suitability for real-time deployment. These limitations motivate a unified approach that integrates semantic reasoning with efficient, adversary-aware consensus tailored to smart-city security.

The literature shows that existing work lacks a unified mechanism for semantic reasoning, dynamic consensus formation, and adversary-aware decision-making in smart cities. Prior studies address optimization, detection accuracy, and multi-agent collaboration separately, but they do not explain how distributed evidence should be interpreted, validated, and aggregated under adversarial conditions. Classical swarm intelligence methods, such as ACO, provide structured mechanisms for distributed coordination and numerical optimization, but they are not designed to reason over semantic, conflicting, and context-dependent evidence. To address this gap, we propose a swarm-driven multi-agent reasoning approach that integrates semantic intelligence with the TPSC mechanism. This approach enables dynamic belief aggregation, contradiction-aware reasoning, and robust decision-making under uncertainty in smart-city environments.

\section{Threat Model}
\label{Threat Model}
TPSC-Sec considers a heterogeneous smart city in which sensing devices, actuators, gateways, edge nodes, and cloud services continuously exchange telemetry and control messages. The adversary can affect both the observable cyber-physical environment and the internal multi-agent reasoning process by introducing noisy, incomplete, and misleading evidence. The environment is modeled as a partially observed sequential process over time windows \(t=1,\ldots,T\), where the true security state evolves through stages such as benign operation, reconnaissance, persistence, lateral movement, and command-and-control activity. The defender observes only partial and potentially manipulated evidence:
\[
E_t \sim \mathbb{P}(E_t \mid s_t, a_t),
\]
where \(s_t\) denotes the latent security state and \(a_t\) denotes the adversarial action. This formulation captures decision-making under incomplete and adversarial observations.

\subsection{Environment and Protected Assets}
The system is represented as a time-varying graph \(G_t=(V_t,\mathcal{E}_t)\), where nodes correspond to smart-city entities and edges represent communication relationships. Node attributes capture behavioral, protocol, historical, and functional properties, while edges describe interaction dynamics. To connect this graph representation to security objectives, we define the protected security dimensions as:
\[
\mathcal{A}_{\mathrm{sec}} = \{a_{\mathrm{beh}}, a_{\mathrm{comm}}, a_{\mathrm{proto}}, a_{\mathrm{id}}, a_{\mathrm{avail}}\},
\]
corresponding to behavioral integrity, communication legitimacy, protocol correctness, identity consistency, and availability. These dimensions represent the main assets that TPSC-Sec aims to monitor and protect. The global system state is defined as:
\[
\mathbf{z}_t = [z_{\mathrm{beh}}, z_{\mathrm{comm}}, z_{\mathrm{proto}}, z_{\mathrm{id}}, z_{\mathrm{avail}}]^\top,
\quad
\mathcal{S}_{\mathrm{global}}(t) = \sum_q \omega_q z_q(t),
\]
where \(z_q(t)\) denotes the security status of dimension \(q\), and \(\omega_q\) represents its operational importance. This provides a compact and interpretable representation of the overall security condition of the smart-city system.

\subsection{Adversary Model}
The adversary is adaptive and can operate externally, from semi-trusted positions, and through compromised nodes, with access to node and edge sets \(V_t^{\mathcal{A}}\) and \(E_t^{\mathcal{A}}\). It performs multi-stage attacks defined as:
\[
\Gamma_{\mathcal{A}} = (c_{t_1},\ldots,c_{t_n}),
\]
including reconnaissance, spoofing, injection, persistence, lateral movement, and denial-of-service. The adversary aims to maximize impact while minimizing detectability:
\[
\max_{\{a_t\}} \sum_{t=1}^{T} \left( \mathcal{I}(a_t) - \lambda_{\mathcal{A}} \mathcal{D}(a_t) \right),
\]
where detectability depends on hypothesis separation:
\[
\mathcal{D}(a_t) = \mathbb{E}\!\left[P_t(h^{+}) - \max_{h \neq h^{+}} P_t(h)\right]_{+}.
\]

\subsection{Attack Surface}
The attack surface spans four channels: i. traffic behavior, ii. protocol interaction, iii. identity usage, and iv. temporal dynamics. Attacks often appear as distributed, low-intensity signals, in which individual indicators remain below the detection thresholds but collectively exceed a global threshold. This formulation captures stealthy multi-stage behavior that evades local detectors but becomes detectable through integrated reasoning.

\subsection{Impact on Reasoning and Consensus}
The adversary also targets collective reasoning by increasing belief fragmentation:
\[
\mathcal{F}_t = 1 - \frac{P_t(h^{+})}{\sum_{h} P_t(h) + \varepsilon},
\]
and reducing the consensus margin:
\[
\bar{\Delta}_t =
\frac{P_t(h^{+}) - \max_{h \neq h^{+}} P_t(h)}
{\sum_{h} P_t(h) + \varepsilon}.
\]
The adversarial objective is therefore defined as:
\[
\min_{\{a_t\}} \sum_{t=1}^{T} \bar{\Delta}_t,
\]
which weakens decision confidence and increases ambiguity in the reasoning process.

\subsection{Assumptions and Scope}
TPSC-Sec assumes a partially trusted architecture, in which the coordination, verification, and control modules are trusted, whereas external entities and compromised nodes are not. Evidence quality is bounded:
\[
0 < \varphi_{\min} \le \varphi_t \le 1,
\]
allowing noisy and adversarial observations. The threat model includes reconnaissance, protocol misuse, identity attacks, persistence, lateral movement, and multi-stage campaigns. Physical attacks and compromise of trusted components are out of scope.

\begin{figure*}[ht]
\centering 
\includegraphics[width=0.83\linewidth]{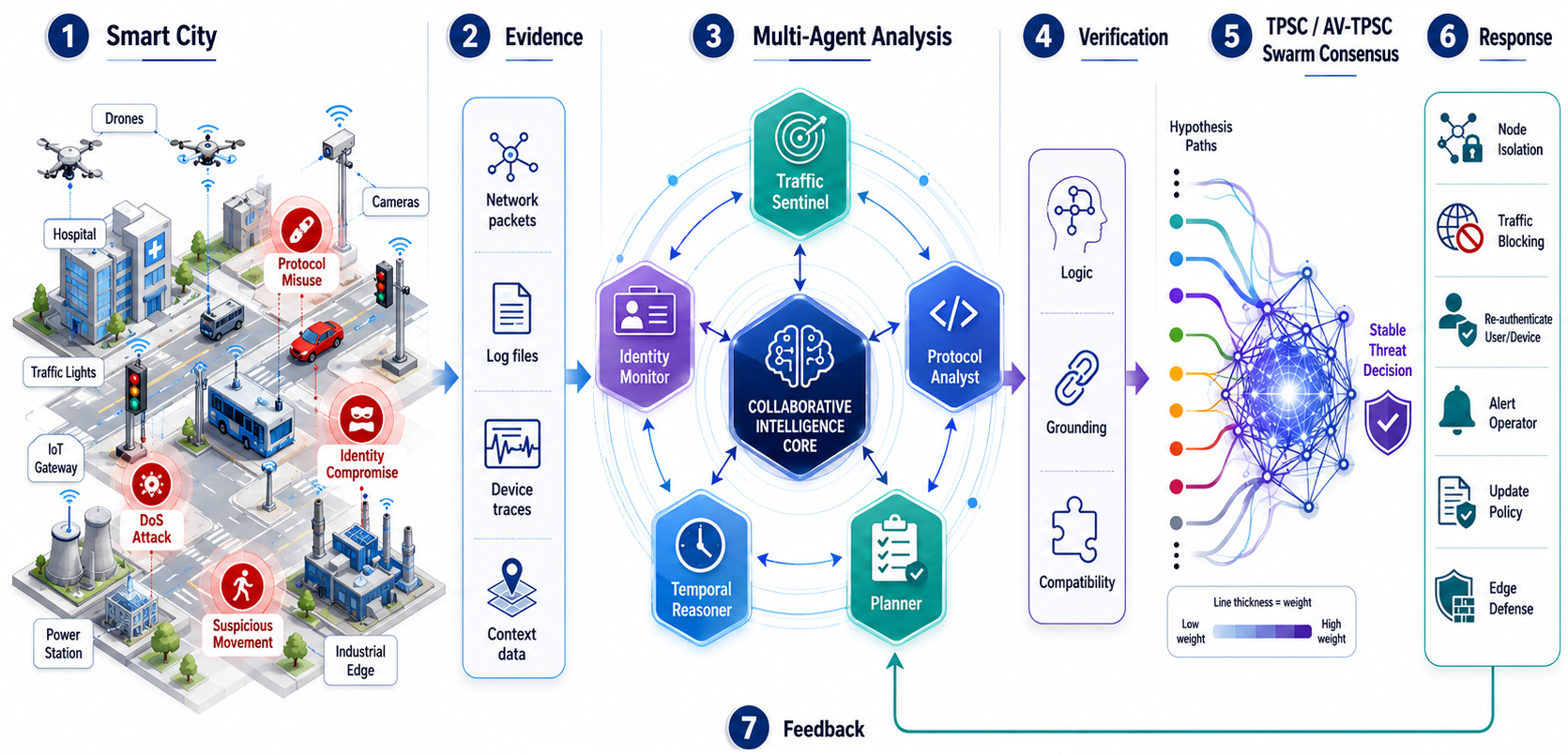} 
\caption{ Unified architecture of TPSC-Sec integrating smart city monitoring, multi-agent LLM reasoning, verification, and TPSC for belief aggregation. Adversarial impact across traffic, protocol, identity, and temporal attack surfaces affects both observed data and agent-level reasoning dynamics. } 
\label{fig:tpsc_architecture} 
\end{figure*}

\section{Methodology}
\label{Methodology}
This section presents \textit{TPSC-Sec}, a distributed security reasoning solution that integrates a multi-agent LLM with a swarm-driven (ACO) consensus mechanism to enable stable and interpretable decision-making under adversarial uncertainty in a smart city. The central challenge addressed in this work is the instability of distributed reasoning. Single-agent LLMs suffer from hallucination and inconsistency \cite{lin2025llm}\cite{xu2026mitigating}, while centralized approaches do not scale reliably in dynamic smart-city settings. TPSC-Sec addresses this limitation by introducing a coordinated reasoning process in which multiple specialized agents collaboratively construct, validate, and refine threat interpretations through structured interaction. The mechanism decomposes security analysis into semantically distinct reasoning tasks. Each agent operates on a specific evidence projection and produces a structured hypothesis grounded in contextual observations. Instead of treating these outputs independently, TPSC-Sec models collective reasoning as a time-dependent process of belief formation. Hypotheses evolve through a pheromone-based mechanism that reinforces consistent evidence, suppresses contradictions, and stabilizes consensus over time. The system consists of three tightly integrated components: 1) a multi-agent reasoning layer that generates semantically grounded hypotheses, 2) a TPSC module that governs belief dynamics, and 3) an edge-level decision module that maps validated threat states to near-optimal mitigation actions. These components operate as a unified solution in which reasoning, verification, and decision-making are tightly coupled, as illustrated in Figure~\ref{fig:tpsc_architecture}. Formally, the solution evolves over observation windows \(t = 1, 2, \ldots, T\). At each step, telemetry is transformed into structured, contextual evidence, processed by specialized agents, verified for logical consistency and grounding, and aggregated via swarm dynamics to produce a stable belief about threat hypotheses. The resulting belief directly drives the selection of a near-optimal response action. The global inference process is defined as:
\[
E_t \xrightarrow{\text{Agents}} \mathcal{H}_t
\xrightarrow{\text{Verification}} \hat{\mathcal{H}}_t
\xrightarrow{\text{TPSC}} P_{t+1}(h)
\xrightarrow{\text{Decision}} r_t^* .
\]
Moreover, TPSC-Sec aims to maximize decision reliability while controlling false alarms, latency, and operational cost:
\[
\max_{\Pi}
\sum_{t=1}^{T}
\left(
\lambda_1 \mathrm{TPR}_t
-
\lambda_2 \mathrm{FPR}_t
-
\lambda_3 \mathrm{Delay}_t
-
\lambda_4 \mathrm{Cost}_t
\right),
\]
where \(\Pi\) denotes the global decision policy induced by the coupled operation of reasoning, verification, and swarm consensus. To explicitly model uncertainty, the system is formulated as a partially observed sequential process in which the true security state remains latent and must be inferred from incomplete and noisy evidence:
\[
E_t \sim \mathbb{P}(E_t \mid s_t), \quad
s_{t+1} \sim \mathbb{P}(s_{t+1} \mid s_t, r_t^*).
\]
This formulation establishes TPSC-Sec as a unified approach in which security is modeled as a dynamic, collective evolution of beliefs, enabling robust, adversary-resilient reasoning beyond isolated prediction mechanisms.

\subsection{Multi-Agent LLM Security Analysts}
The multi-agent reasoning layer performs distributed security analysis by decomposing the overall reasoning process into semantically distinct tasks. Each agent processes smart-city telemetry and produces structured analytical insights aligned with its designated reasoning domain. This design follows the principle that heterogeneous evidence requires specialized inference rather than uniform processing. Formally, let the agent set be defined as:
\[
A = \{A_1,A_2,\ldots,A_m\},
\]
where each agent \(A_i\) implements a role-specific reasoning function:
\[
A_i : (E_i,C_t) \mapsto \tilde{h}_i,
\]
with \(E_i = \Gamma_i(E_t)\) denoting the agent-specific evidence projection, \(C_t\) the contextual system state, and \(\tilde{h}_i\) the resulting structured hypothesis. The overall reasoning process is defined as:
\[
\mathfrak{A}(E_t,C_t) = \{A_1(E_1,C_t),\ldots,A_m(E_m,C_t)\}.
\]
This formulation enforces structured specialization by projecting global evidence:
\[
E_t = (F_t,P_t,D_t,L_t,C_t)
\]
into role-specific subspaces:
\[
E_i = \Gamma_i(E_t), \qquad i=1,\ldots,m,
\]
ensuring that each agent operates on a semantically coherent subset of information. The \textit{Traffic Sentinel} agent analyzes flow-level behavior by identifying deviations in communication dynamics. It evaluates packet rate, connection fan-out, burst patterns, and destination entropy to construct interpretable indicators of abnormal behavior. Instead of producing isolated anomaly scores, it generates structured reasoning outputs. Let \(F_t\) denote the flow-level feature matrix. The resulting representation is defined as:
\[
a_{\mathrm{flow}}(t) = \Psi_{\mathrm{flow}}(F_t,F_t^{\mathrm{ref}}),
\]
with the decomposition:
\[
a_{\mathrm{flow}}(t)
=
\theta_1 \Delta_{\mathrm{rate}}(t)
+
\theta_2 \Delta_{\mathrm{fanout}}(t)
+
\theta_3 \Delta_{\mathrm{burst}}(t)
+
\theta_4 \Delta_{\mathrm{entropy}}(t).
\]
The entropy term is given by:
\[
H_t = - \sum_{j=1}^{n_t} p_j(t)\log p_j(t),
\]
where deviations indicate abnormal dispersion in communication patterns. The \textit{Protocol Analyst} agent performs semantic evaluation of application-layer interactions. It analyzes protocol sequences and command structures to identify inconsistencies between observed behavior and expected constraints, enabling detection of semantically invalid actions that remain syntactically correct. Let \(P_t = (p_1,p_2,\ldots,p_\ell)\) denote the protocol sequence. The inconsistency score is defined as:
\[
a_{\mathrm{proto}}(t) = \Psi_{\mathrm{proto}}(P_t,\Pi_{\mathrm{role}},\Pi_{\mathrm{state}}),
\]
with:
\[
a_{\mathrm{proto}}(t)
=
\mu_1 \chi_{\mathrm{cmd}}(t)
+
\mu_2 \chi_{\mathrm{path}}(t)
+
\mu_3 \chi_{\mathrm{state}}(t)
+
\mu_4 \chi_{\mathrm{priv}}(t),
\]
where the transition violation term is:
\[
\chi_{\mathrm{state}}(t)
=
\sum_{\tau=1}^{\ell-1}
\mathbf{1}\big[(p_{\tau},p_{\tau+1}) \notin \mathcal{T}_{\mathrm{proto}}\big].
\]
The \textit{Identity Monitor} agent analyzes identity usage and behavioral consistency across entities. It detects impersonation, spoofing, and role violations by comparing observed activity against historical patterns and expected profiles, capturing inconsistencies that are not visible at the protocol and flow level. The \textit{Temporal Reasoner} models dependencies across observation windows to capture sequential attack patterns. It identifies multi-stage attack progression, including reconnaissance, persistence, and lateral movement, which cannot be detected from single-step observations. The \textit{Planner} agent integrates outputs from all reasoning agents and combines them with contextual system information. It produces a global structured hypothesis that is consistent across agents and aligned with system objectives, ensuring that decisions are context-aware and operationally actionable. 

\subsection{Threat-Pheromone Swarm Consensus}
The proposed TPSC mechanism is inspired by pheromone-based collective behavior in swarm intelligence, particularly ACO, while extending these principles from numerical optimization to semantic belief aggregation in multi-agent reasoning systems. TPSC aggregates beliefs across hypotheses generated by distributed agents through a pheromone-driven consensus mechanism. The system models collective reasoning as a time-dependent process of belief evolution, where semantic hypotheses are reinforced, inhibited, and stabilized across the agent population. Let \( h \in \mathcal{H}_t \) denote a candidate threat hypothesis produced by the multi-agent reasoning process. Each hypothesis represents a semantically grounded interpretation of system behavior. Each agent emits a pheromone value reflecting the strength of its hypothesis:
\[
\phi_i(h) = c_i \cdot e_i
\]
This formulation ensures that each hypothesis's contribution depends on both confidence and supporting evidence, preventing the dominance of weak, poorly grounded interpretations. The aggregated pheromone signal is defined as:
\[
\Phi(h) = \sum_{i=1}^{m} w_i \phi_i(h),
\]
where \(w_i\) denotes agent reliability weights derived from historical performance and contextual relevance.
To incorporate urgency, pheromone emission is extended as:
\[
\phi_i^{+}(h) = w_i\, c_i^{\alpha_c} e_i^{\alpha_e} u_i^{\alpha_u},
\]
where \(\alpha_c,\alpha_e,\alpha_u \ge 0\) control the relative impact of confidence, evidence strength, and urgency.
In contrast to classical pheromone models that rely on positive reinforcement, TPSC introduces explicit inhibitory signals to model disagreement among agents. Contradiction is captured through:
\[
\psi_j(h) = \lambda (1-c_j),
\]
which suppresses unsupported hypotheses and prevents premature consensus formation. The resulting consensus score is defined as:
\[
S(h) = \Phi(h) - \sum_j \psi_j(h),
\]
and a hypothesis is accepted when \(S(h) > \tau\). To ensure temporal stability, pheromone states evolve as:
\[
P_{t+1}(h) = (1-\rho)P_t(h) + \eta \Phi_t(h) - \mu \Psi_t(h),
\]
where \(\rho\) controls evaporation, \(\eta\) governs reinforcement, and \(\mu\) regulates contradiction penalties. This dynamic prevents persistence of outdated beliefs and stabilizes consistent hypotheses over time. Intuitively, hypotheses with consistent support accumulate belief, while conflicting and weakly supported hypotheses are attenuated. For interpretability, the pheromone state is normalized as:
\[
\tilde{P}_t(h) = \frac{P_t(h)}{\sum_{h' \in \mathcal{H}_t} P_t(h') + \varepsilon},
\]
yielding a probabilistic interpretation of swarm belief. A softmax-based formulation further refines the belief distribution:
\[
B_t(h)
=
\frac{\exp\left(\zeta P_t(h)\right)}
{\sum_{h' \in \mathcal{H}_t}\exp\left(\zeta P_t(h')\right)},
\]
where \(\zeta\) controls concentration. Let \(\mathcal{H}_t = \{h^{(1)},\ldots,h^{(q)}\}\) denote candidate hypotheses. The selected hypothesis is:
\[
h_t^{*} = \arg\max_{h \in \mathcal{H}_t} P_t(h),
\]
subject to:
\[
P_t(h_t^{*}) > \tau_t.
\]
The adaptive threshold is defined as:
\[
\tau_t = \tau_0 + \beta_1 \mathrm{FPR}_{t-1} + \beta_2 \mathrm{Load}_t + \beta_3 \mathrm{Crit}_t,
\]
allowing sensitivity to adjust based on system conditions. Consensus stability is quantified by the margin:
\[
\Delta_t
=
P_t(h_t^{*})
-
\max_{h \neq h_t^{*}} P_t(h),
\]
with normalized form:
\[
\bar{\Delta}_t
=
\frac{P_t(h_t^{*}) - \max_{h \neq h_t^{*}} P_t(h)}
{\sum_{h \in \mathcal{H}_t} P_t(h)+\varepsilon}.
\]
This formulation establishes TPSC as a belief-driven consensus mechanism that stabilizes distributed reasoning under uncertainty through reinforcement, contradiction, and temporal dynamics.

\subsection{Adaptive Verified Threat-Pheromone Swarm Consensus}
The \textit{AV-TPSC} extends the \textit{TPSC} mechanism by introducing verification-aware belief modulation, context-sensitive reinforcement, and adaptive contradiction control to ensure stable reasoning under uncertainty and adversarial conditions. The core principle of AV-TPSC is that hypothesis reliability evolves dynamically with respect to verification quality, contextual relevance, and inter-agent agreement. Accordingly, the consensus process is augmented to directly incorporate these factors into the pheromone dynamics. First, each agent’s pheromone contribution is adjusted using verification-aware confidence. The positive pheromone emission is defined as:
\[
\phi_i^{\mathrm{AV}}(h)
=
w_i \,
(c_i^{\mathrm{ver}})^{\alpha_c}
(\hat{e}_i)^{\alpha_e}
u_i^{\alpha_u},
\]
where \(c_i^{\mathrm{ver}} = \hat{c}_i v_i\) combines calibrated confidence with verification quality. This formulation attenuates hypotheses that lack logical grounding and are inconsistent with peer reasoning, ensuring that belief reinforcement remains evidence-consistent. Second, the consensus mechanism incorporates context-sensitive modulation through a function \(\gamma(C_t)\), which encodes system-level importance and operational risk. The aggregated pheromone is defined as:
\[
\Phi_t^{\mathrm{AV}}(h)
=
\gamma(C_t)
\sum_{i=1}^{m}
\phi_i^{\mathrm{AV}}(h).
\]
This modulation amplifies belief formation in high-criticality contexts, enabling the system to prioritize threats affecting sensitive infrastructure. Third, the dynamics of contradiction are governed by an adaptive scaling mechanism. The inhibitory term is defined as:
\[
\Psi_t^{\mathrm{AV}}(h)
=
\sum_{j=1}^{m}
\lambda_t (1-c_j^{\mathrm{ver}})
\mathbf{1}[y_j \neq h],
\]
where the contradiction sensitivity parameter evolves as:
\[
\lambda_t
=
\lambda_0
\left(
1 +
\kappa \mathcal{D}_{\mathrm{ens}}
\right).
\]
This formulation increases the penalty for contradictions when disagreement is high, suppressing unstable hypotheses and preventing fragmented reasoning from dominating the swarm state. The pheromone update dynamics are therefore expressed as:
\[
P_{t+1}(h)
=
(1-\rho)P_t(h)
+
\eta \Phi_t^{\mathrm{AV}}(h)
-
\mu \Psi_t^{\mathrm{AV}}(h).
\]
Intuitively, AV-TPSC strengthens hypotheses that are both well-supported and well-verified, while dynamically suppressing unreliable and inconsistent interpretations when disagreement is high. It integrates three tightly coupled mechanisms: 1) verification-aware pheromone emission, 2) context-sensitive belief reinforcement, and 3) disagreement-adaptive contradiction scaling. These mechanisms collectively stabilize belief evolution under noisy and adversarial conditions, reduce the propagation of hallucinations, and align consensus formation with system-level priorities. As a result, AV-TPSC enables reliable and interpretable collective reasoning in a large-scale smart city.

\subsection{Multi-Agent Threat Reasoning Algorithm}
The multi-agent reasoning stage transforms structured telemetry into semantically grounded hypotheses through coordinated and role-specialized inference. Each agent operates within the same observation window but from a distinct analytical perspective, enabling diverse yet structured interpretations that support stable consensus.
\begin{algorithm}[t]
\caption{Multi-Agent Threat Reasoning}
\label{alg:reasoning}
\small
\begin{algorithmic}[1]
\Require Observation window $E_t$, agent set $A = \{A_1,\ldots,A_m\}$
\Ensure Hypothesis set $\mathcal{H}_t$

\State Initialize $\mathcal{H}_t \gets \emptyset$

\For{each agent $A_i \in A$}
    \State Extract agent-specific evidence slice: $E_i \gets \Gamma_i(E_t)$
    \State Construct contextual prompt: $Q_i \gets \Pi_i(E_i, C_t)$
    \State Generate reasoning output: $z_i \gets \mathrm{LLM}_i(Q_i)$
    \State Infer threat label: $y_i \gets \Upsilon_i(z_i)$
    \State Estimate confidence: $c_i \gets \mathrm{Conf}(z_i, E_i)$
    \State Estimate evidence strength: $e_i \gets \mathrm{Evid}(E_i)$
    \State Estimate urgency: $u_i \gets \mathrm{Urg}(y_i, C_t)$
    \State Propose response: $r_i \gets \mathrm{Resp}(y_i, u_i, C_t)$
    \State Form hypothesis: $\tilde{h}_i \gets (y_i, c_i, e_i, u_i, r_i, z_i)$
    \State Update hypothesis set: $\mathcal{H}_t \gets \mathcal{H}_t \cup \{\tilde{h}_i\}$
\EndFor

\State \textbf{return} $\mathcal{H}_t$
\end{algorithmic}
\end{algorithm}
Algorithm~\ref{alg:reasoning} formalizes the distributed reasoning process, where each agent operates on a role-specific evidence projection to ensure semantic consistency. The contextual prompt integrates system state \(C_t\), grounding inference, and reducing inconsistency across agents. The functions \(\mathrm{Conf}(\cdot)\), \(\mathrm{Evid}(\cdot)\), and \(\mathrm{Urg}(\cdot)\) estimate confidence, evidence strength, and urgency through calibrated scoring, preserving interpretability. Each agent maps its contextualized evidence to a structured hypothesis, and the resulting hypotheses are aggregated into a unified set \(\mathcal{H}_t\). This formulation separates evidence projection, reasoning, and hypothesis construction, enabling controlled specialization while maintaining global consistency. The computational cost scales linearly with the number of agents, providing a clear characterization of system scalability.

\subsection{Swarm Consensus Algorithm}
The swarm consensus stage transforms the hypothesis set into a stable collective decision through pheromone-driven belief dynamics. Inspired by pheromone-based collective behavior in swarm intelligence, particularly ACO, this mechanism resolves conflicting agent outputs by integrating reinforcement, contradiction inhibition, and reliability-aware aggregation within a unified update process.
\begin{algorithm}[t]
\caption{Threat-Pheromone Swarm Consensus}
\label{alg:tpsc}
\small
\begin{algorithmic}[1]
\Require Hypothesis set $\mathcal{H}_t$, weights $W$, previous state $P_t(h)$
\Ensure Updated state $P_{t+1}(h)$, accepted hypothesis $h_t^*$

\For{each hypothesis $\tilde{h}_i \in \mathcal{H}_t$}
    \State Extract $(y_i,c_i,e_i,u_i,r_i,z_i)$
    \State $\phi_i \gets c_i \cdot e_i$
    \State $\Phi_t(y_i) \gets \Phi_t(y_i) + w_i \phi_i$
\EndFor

\For{each label $h \in \mathcal{Y}_{\text{threat}}$}
    \State $\Psi_t(h) \gets 0$
    \For{each $\tilde{h}_j \in \mathcal{H}_t$}
        \If{$y_j \neq h$}
            \State $\Psi_t(h) \gets \Psi_t(h) + \lambda (1-c_j)$
        \EndIf
    \EndFor
    \State $P_{t+1}(h) \gets (1-\rho)P_t(h) + \eta \Phi_t(h) - \mu \Psi_t(h)$
\EndFor

\State $h_t^* \gets \arg\max_{h} P_{t+1}(h)$

\If{$P_{t+1}(h_t^*) > \tau_t$}
    \State Accept $h_t^*$
\Else
    \State Unresolved/benign
\EndIf

\State \textbf{return} $P_{t+1}(h), h_t^*$
\end{algorithmic}
\end{algorithm}
Algorithm~\ref{alg:tpsc} defines a sequential belief update process in which pheromone signals accumulate support for competing hypotheses while suppressing inconsistent alternatives. These dynamics enable stable consensus formation by reinforcing coherent interpretations and attenuating the impact of conflicting evidence over time. This process can be interpreted as a belief-driven extension of classical swarm mechanisms, such as ACO, to semantic reasoning. Agent reliability is updated online based on validation outcomes, ensuring bounded and comparable impact across agents. This allows agents to contribute consistently to the consensus process while maintaining stability over time.

\subsection{Autonomous Response}
The response stage maps the stabilized swarm belief to an executable mitigation action. The decision process is tightly coupled with the consensus output, ensuring that selected actions reflect both validated threat semantics and system context. Let:
\[
R = \{r_1,r_2,\ldots,r_k\}
\]
denote the set of candidate mitigation actions. Each action is evaluated using a cost function:
\[
C(r) = \delta_1 D(r) + \delta_2 L(r) + \delta_3 O(r),
\]
where \(D(r)\) represents residual damage, \(L(r)\) deployment latency, and \(O(r)\) operational overhead. The near-optimal action is selected as:
\[
r^* = \arg\min_{r\in R} C(r).
\]
This formulation prioritizes actions that minimize system impact while maintaining low latency and resource cost. The cost components capture the trade-off between mitigation effectiveness and operational disruption. To account for uncertainty in the accepted hypothesis, the decision process is extended using an expected utility formulation:
\[
U(r \mid h_t^*) = P(h_t^*)\, B(r,h_t^*) - C(r),
\]
where \(P(h_t^*)\) denotes swarm confidence and \(B(r,h_t^*)\) the expected mitigation benefit. The selected action satisfies:
\[
r^* = \arg\max_{r \in R_f} U(r \mid h_t^*),
\]
where \(R_f \subseteq R\) is the feasible action set constrained by system policy and operational context.

\subsection{Agent Verification Procedure}
The verification stage regulates the quality of hypotheses prior to swarm aggregation, ensuring that belief formation remains grounded, consistent, and stable under uncertainty. It acts as a control layer that attenuates unreliable reasoning outputs before they impact pheromone dynamics.
\begin{algorithm}[t]
\caption{Hypothesis Verification}
\label{alg:verify}
\small
\begin{algorithmic}[1]
\Require Hypothesis set $\mathcal{H}_t$
\Ensure Verified set $\hat{\mathcal{H}}_t$

\State $\hat{\mathcal{H}}_t \gets \emptyset$

\For{$\tilde{h}_i = (y_i,c_i,e_i,u_i,r_i,z_i) \in \mathcal{H}_t$}
    \State $\ell_i \gets \mathrm{Logic}(z_i)$
    \State $g_i \gets \mathrm{Ground}(z_i, E_i)$
    \State $q_i \gets \mathrm{Compat}(\tilde{h}_i, \mathcal{H}_t \setminus \{\tilde{h}_i\})$
    \State $v_i \gets \xi_1 \ell_i + \xi_2 g_i + \xi_3 q_i$
    \If{$v_i > \zeta$}
        \State Add $\tilde{h}_i$ to $\hat{\mathcal{H}}_t$
    \Else
        \State $c_i \gets \nu c_i$
        \State Add adjusted $\tilde{h}_i$ to $\hat{\mathcal{H}}_t$
    \EndIf
\EndFor

\State \Return $\hat{\mathcal{H}}_t$
\end{algorithmic}
\end{algorithm}
Algorithm~\ref{alg:verify} formalizes verification as a structured evaluation process. Each hypothesis is assessed along three dimensions: logical consistency, evidence grounding, and cross-agent compatibility. A combined verification score determines the effective contribution of each hypothesis to the reasoning process. Rather than discarding inconsistent outputs, the procedure attenuates their impact through confidence scaling. This design preserves informative signals while preventing unstable and weakly grounded reasoning from dominating belief aggregation.
The aggregate verification strength provides a global measure of hypothesis quality at each time step. Lower values indicate reduced consistency and insufficient grounding, which can be used to adjust downstream decision thresholds. The adjusted confidence propagated to the consensus stage ensures that hypotheses with lower verification quality exert less impact on pheromone emission.

\subsection{Global Mathematical Representation of TPSC-Sec}
TPSC-Sec is formulated as a stochastic multi-agent decision process operating over time-indexed evidence streams in a heterogeneous smart city. The system state at time \(t\) is defined as:
\[
\mathcal{S}_t =
(E_t,\mathcal{H}_t,P_t,W_t,C_t),
\]
where \(E_t\) denotes structured evidence, \(\mathcal{H}_t\) the hypothesis set, \(P_t\) the pheromone state over hypotheses, \(W_t\) the agent reliability weights, and \(C_t\) the contextual system information. The global state transition is defined as:
\[
\mathcal{S}_{t+1}
=
\mathcal{F}(\mathcal{S}_t),
\]
where \(\mathcal{F}\) captures the coupled operations of reasoning, verification, swarm consensus, and response. This transition can be decomposed as:
\[
\mathcal{S}_{t+1}
=
\left(
E_{t+1},
\mathfrak{A}(E_t,C_t),
\mathfrak{T}(P_t,\hat{\mathcal{H}}_t),
\mathfrak{W}(W_t),
C_{t+1}
\right),
\]
where \(\mathfrak{A}(\cdot)\) denotes multi-agent reasoning, \(\mathfrak{T}(\cdot)\) pheromone update dynamics, and \(\mathfrak{W}(\cdot)\) reliability adaptation. Let the hypothesis space be:
\[
\mathcal{H} =
\{h^{(1)},h^{(2)},\ldots,h^{(q)}\}.
\]
The normalized belief distribution is:
\[
B_t(h)
=
\frac{P_t(h)}
{\sum_{h' \in \mathcal{H}} P_t(h') + \varepsilon},
\]
providing a probabilistic interpretation of swarm belief. The selected hypothesis is:
\[
h_t^{*}
=
\arg\max_{h} B_t(h).
\]
Consensus stability is quantified by the margin:
\[
\Delta_t
=
B_t(h_t^{*})
-
\max_{h \neq h_t^{*}} B_t(h),
\]
where larger values indicate stronger agreement across agents. Over a horizon \(T\), detection performance is defined as:
\[
\mathbb{E}[\mathrm{TPR}]
=
\frac{1}{T}
\sum_{t=1}^{T}
\mathbf{1}[h_t^{*}=h_t^{+}],
\]
and the false-positive rate as:
\[
\mathrm{FPR}
=
\frac{1}{T}
\sum_{t=1}^{T}
\mathbf{1}[h_t^{*}\neq h_t^{+} \land h_t^{*}\neq y_{\mathrm{benign}}].
\]
Let \(r_t^*\) denote the selected mitigation action. The cumulative operational cost is:
\[
\mathcal{J}
=
\sum_{t=1}^{T}
\Big(
\alpha_1 D(r_t^*)
+
\alpha_2 L(r_t^*)
+
\alpha_3 O(r_t^*)
\Big).
\]

\section{Experimental Setup}
\label{Experimental Setup}
TPSC-Sec is implemented as a multi-agent LLM-based system in which specialized agents perform traffic analysis, protocol inspection, temporal reasoning, verification, and response planning. Agents operate concurrently and exchange structured hypotheses through the \textit{TPSC} mechanism. Telemetry is processed using a fixed-size sliding-window scheme, where each window is mapped to a structured evidence object \(E_t\). All experiments are conducted over 500 independent runs to ensure statistical reliability. We use the ToN-IoT dataset \cite{toniot_dataset_unsw} as a representative smart-city benchmark that contains network flows, protocol traces, and device-level activity. Raw samples are transformed into structured evidence:
\[
E_t = (F_t, P_t, D_t, L_t, C_t),
\]
where \(F_t\), \(P_t\), \(D_t\), \(L_t\), and \(C_t\) denote flow statistics, protocol behavior, device attributes, system logs, and contextual information, respectively. Each agent is implemented using the LLM with role-specific prompting. A low temperature setting (0.2) is used to reduce stochastic variability, and outputs are constrained to structured formats including threat label, confidence, evidence strength, urgency, and response suggestion. All agents operate under identical configurations to ensure consistency. Experiments are conducted on a standard workstation. To ensure fair comparison, we evaluate against two baselines: 1) a \textit{Single-Agent LLM} that processes the full input \(E_t\) without specialization, and 2) a \textit{Multi-Agent Majority Voting} approach, where multiple agents generate hypotheses but aggregation is performed via simple voting without verification. Evaluation follows a scenario-driven protocol in which diverse attack patterns, including reconnaissance, denial-of-service, protocol misuse, and multi-stage attacks, are presented as sequential observation windows. Each window is processed through multi-agent reasoning, verification, and swarm-based consensus. The TPSC parameters \((\rho, \eta, \mu)\) are fixed across all experiments. The verification module enforces logical consistency, evidence grounding, and cross-agent compatibility prior to aggregation. Randomness is controlled through fixed seeds and deterministic configurations. Performance is evaluated using system-level metrics that capture the stability and reliability of distributed reasoning, including consensus margin, aggregate risk, agent agreement, and trust level. These metrics reflect decision stability, safety, and robustness under adversarial conditions.

\section{Experimental Results}
\label{Experimental Results}
This section presents an evaluation of the TPSC-Sec mechanism.

\subsection{Security Risk, Consensus Stability, and Trust Dynamics}
The evaluation shows that TPSC-Sec exhibits stable, controlled security behavior under adversarial and noisy conditions. Results are averaged over 500 runs and reported with the standard deviation, 95\% confidence intervals, and statistical significance.
\begin{figure*}[t]
\centering
\includegraphics[width=0.8\linewidth]{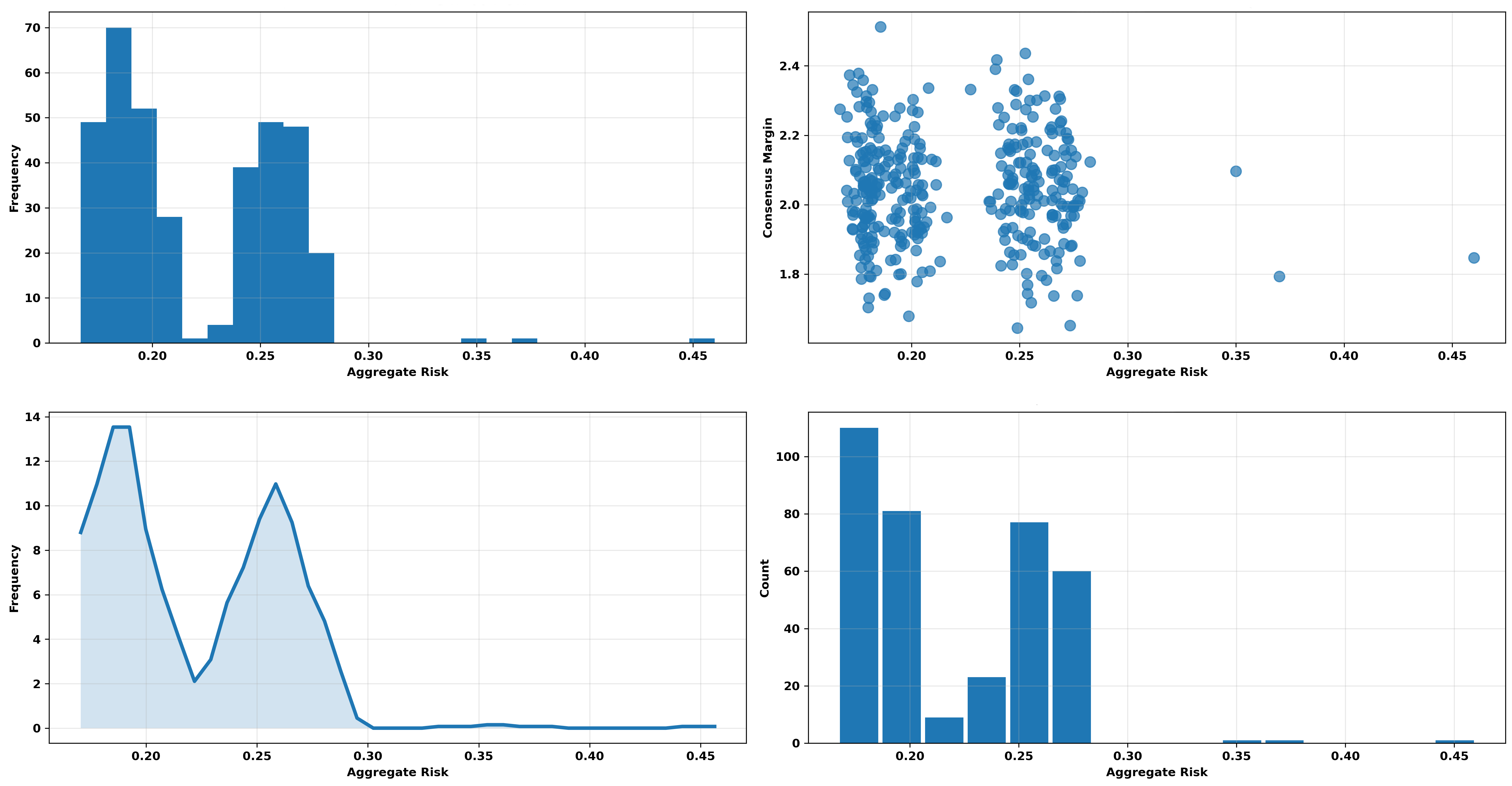}
\caption{Aggregate risk distribution and its relationship with consensus margin across experimental runs.}
\label{fig:aggregate_risk_analysis}
\end{figure*}
Figure~\ref{fig:aggregate_risk_analysis} shows a tightly concentrated aggregate risk distribution with low variance (mean 0.23, std 0.04), indicating effective and consistent risk control. The bimodal structure suggests the presence of two stable operating regimes, while the light upper tail indicates that high-risk states are rare. Importantly, the consensus margin remains positive across runs, ensuring stable separation between competing hypotheses.
\begin{figure*}[t]
\centering
\includegraphics[width=0.80\linewidth]{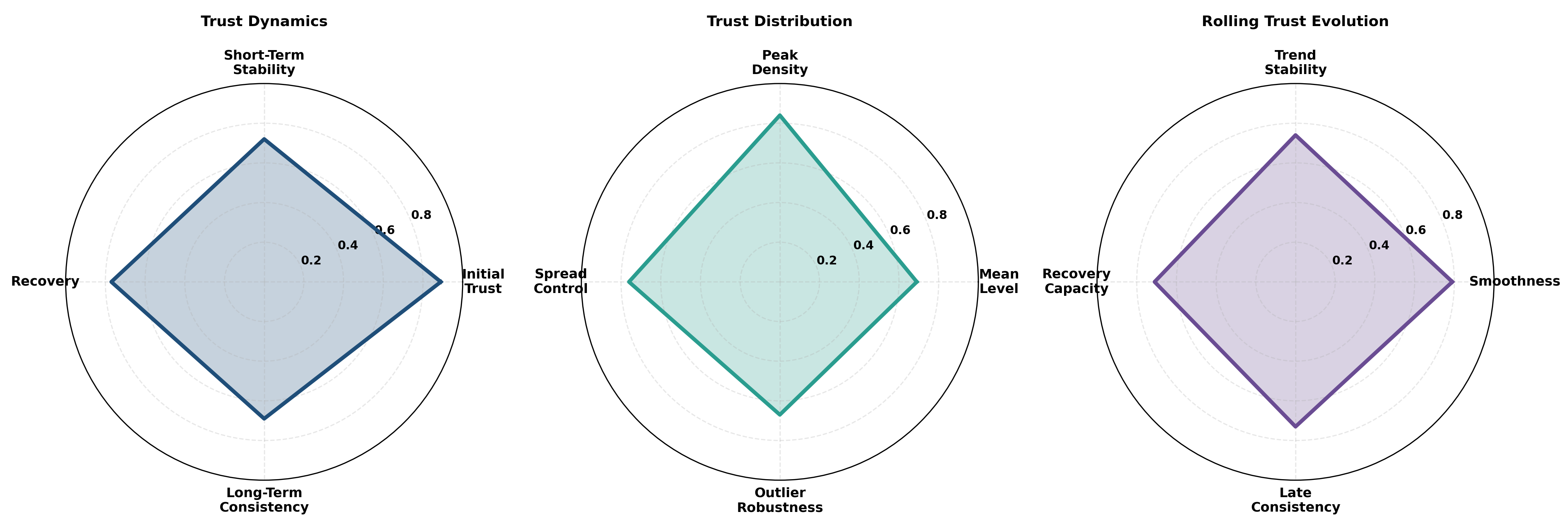}
\caption{Consensus dynamics including belief concentration, margin evolution, and hypothesis alignment.}
\label{fig:consensus_analysis_part1}
\end{figure*}
\begin{figure*}[t]
\centering
\includegraphics[width=0.80\linewidth]{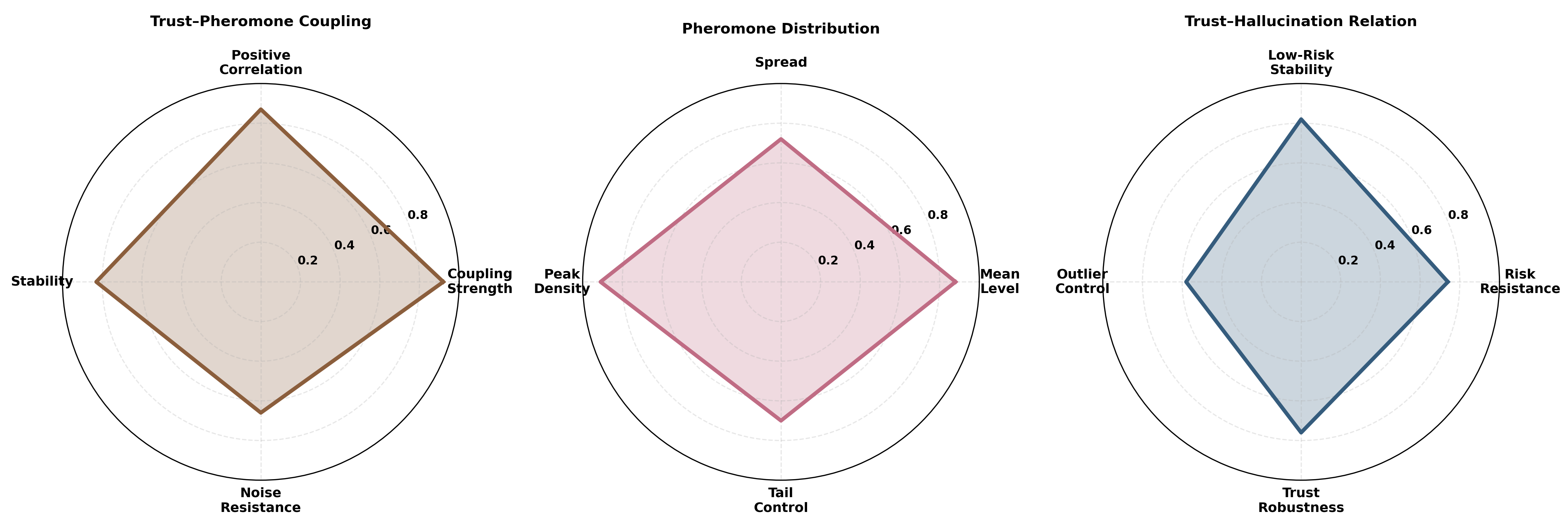}
\caption{System stability analysis showing clustering behavior, variance control, and adaptive thresholding.}
\label{fig:consensus_analysis_part2}
\end{figure*}
Figures~\ref{fig:consensus_analysis_part1} and \ref{fig:consensus_analysis_part2} demonstrate stable consensus formation characterized by monotonic margin growth and controlled dispersion. The observed clustering behavior indicates that system states evolve around stable attractors, avoiding chaotic transitions while preserving responsiveness through adaptive thresholding.
\begin{figure*}[t]
\centering
\includegraphics[width=0.80\linewidth]{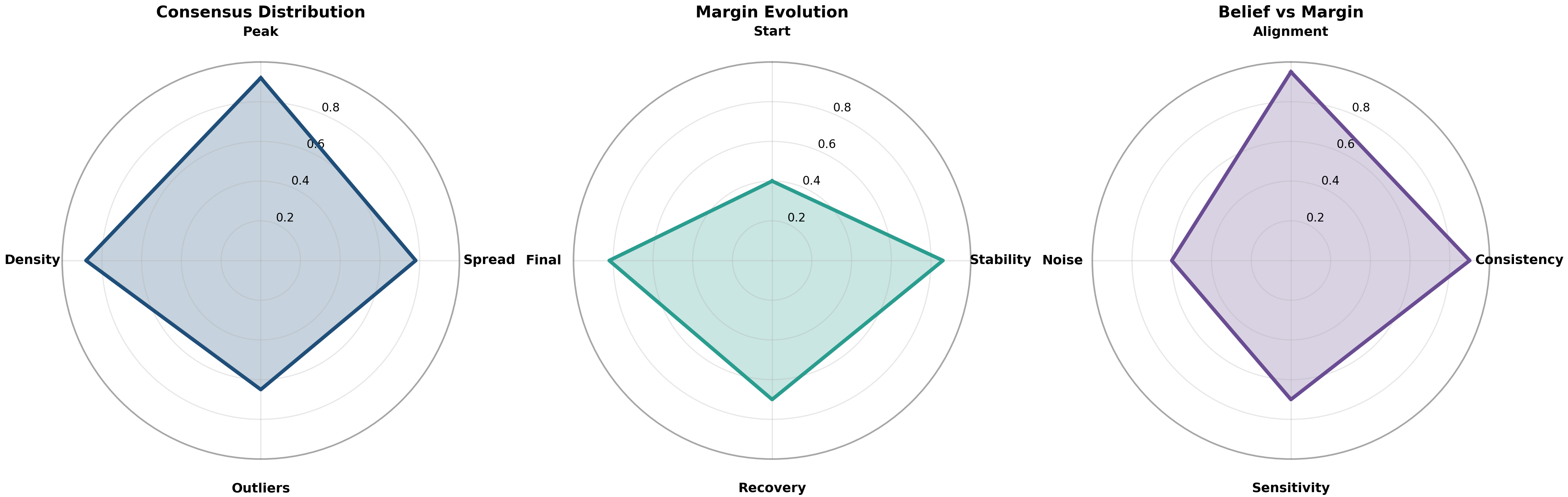}
\caption{Trust evolution across runs showing stability, recovery after disturbances, and long-term consistency.}
\label{fig:trust_analysis}
\end{figure*}
Figure~\ref{fig:trust_analysis} shows stable trust dynamics with rapid recovery following perturbations, indicating a resilient feedback loop between agent performance and trust adaptation.
\begin{figure*}[t]
\centering
\includegraphics[width=0.80\linewidth]{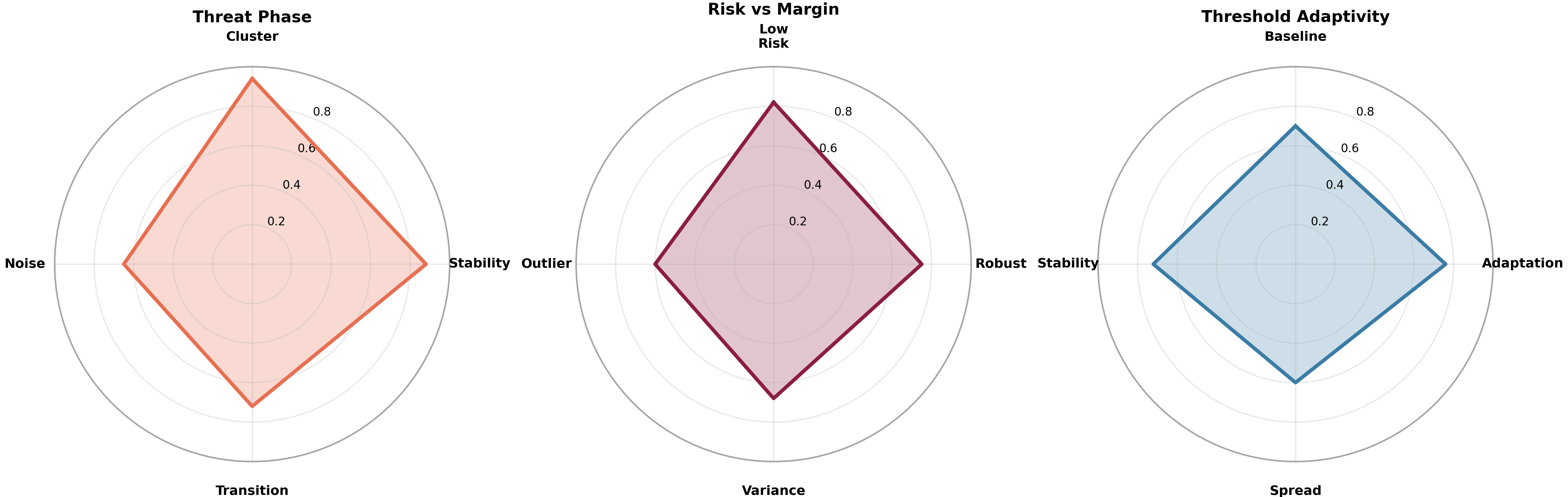}
\caption{Interaction between trust dynamics, pheromone evolution, and system risk.}
\label{fig:trust_pheromone_analysis}
\end{figure*}
Figure~\ref{fig:trust_pheromone_analysis} shows a positive correlation between trust and pheromone reinforcement, leading to improved consensus stability and suppression of unreliable reasoning outputs.
\begin{table}[t]
\centering
\caption{Risk Statistics Across Experimental Runs (with 95\% CI)}
\label{tab:risk_stats}
\renewcommand{\arraystretch}{1.2}
\setlength{\tabcolsep}{4pt}
\resizebox{\linewidth}{!}{
\begin{tabular}{lccccc}
\hline
Metric & Mean & Std & 95\% CI & Min & Max \\
\hline
Aggregate Risk & 0.23 & 0.04 & [0.226, 0.234] & 0.18 & 0.46 \\
Prompt Injection Risk & 0.29 & 0.09 & [0.282, 0.298] & 0.20 & 0.50 \\
Hallucination Propagation Risk & 0.30 & 0.08 & [0.293, 0.307] & 0.20 & 0.50 \\
Tool Misuse Risk & 0.32 & 0.07 & [0.314, 0.326] & 0.20 & 0.48 \\
Data Leakage Risk & 0.27 & 0.05 & [0.266, 0.274] & 0.18 & 0.40 \\
\hline
\end{tabular}
}
\end{table}
Table~\ref{tab:risk_stats} confirms low and stable aggregate risk with statistically significant results ($p < 0.01$). Tool misuse emerges as the dominant risk source (0.32), followed by hallucination propagation (0.30) and prompt injection (0.29), while data leakage has comparatively lower impact.
\begin{table}[t]
\centering
\caption{Effect size and risk contribution analysis}
\label{tab:risk_effect}
\renewcommand{\arraystretch}{1.05}
\setlength{\tabcolsep}{3pt}
\begin{tabular}{lccc}
\hline
\textbf{Interaction} & \textbf{Eff. ($d$)} & \textbf{Role} & \textbf{Interp.} \\
\hline
Tool vs Agg. Risk & 1.35 & Dominant & Actuator/API impact \\
Halluc. vs Agg. Risk & 1.10 & Secondary & Decision distortion \\
Leakage vs Agg. Risk & 0.80 & Moderate & Sensor exposure \\
Prompt Inj. vs Halluc. & -- & Trigger & Adversarial amplification \\
\hline
\end{tabular}
\end{table}
Table~\ref{tab:risk_effect} shows a clear hierarchy of risk contributions ($p < 0.01$), with tool misuse as the dominant factor, followed by hallucination effects, while prompt injection acts as a trigger that amplifies downstream reasoning risks.

\subsection{Multi-Agent Reasoning Performance}
The reasoning subsystem exhibits a clear trade-off between computational cost and decision quality, a trade-off that is critical in a smart city. TPSC-Sec employs five specialized agents, \textit{Traffic Sentinel}, \textit{Protocol Analyst}, \textit{Identity Monitor}, \textit{Temporal Reasoner}, and \textit{Planner}, that process distinct evidence views and collaboratively generate hypotheses for consensus.
\begin{figure}[t]
\centering
\includegraphics[width=0.9\linewidth]{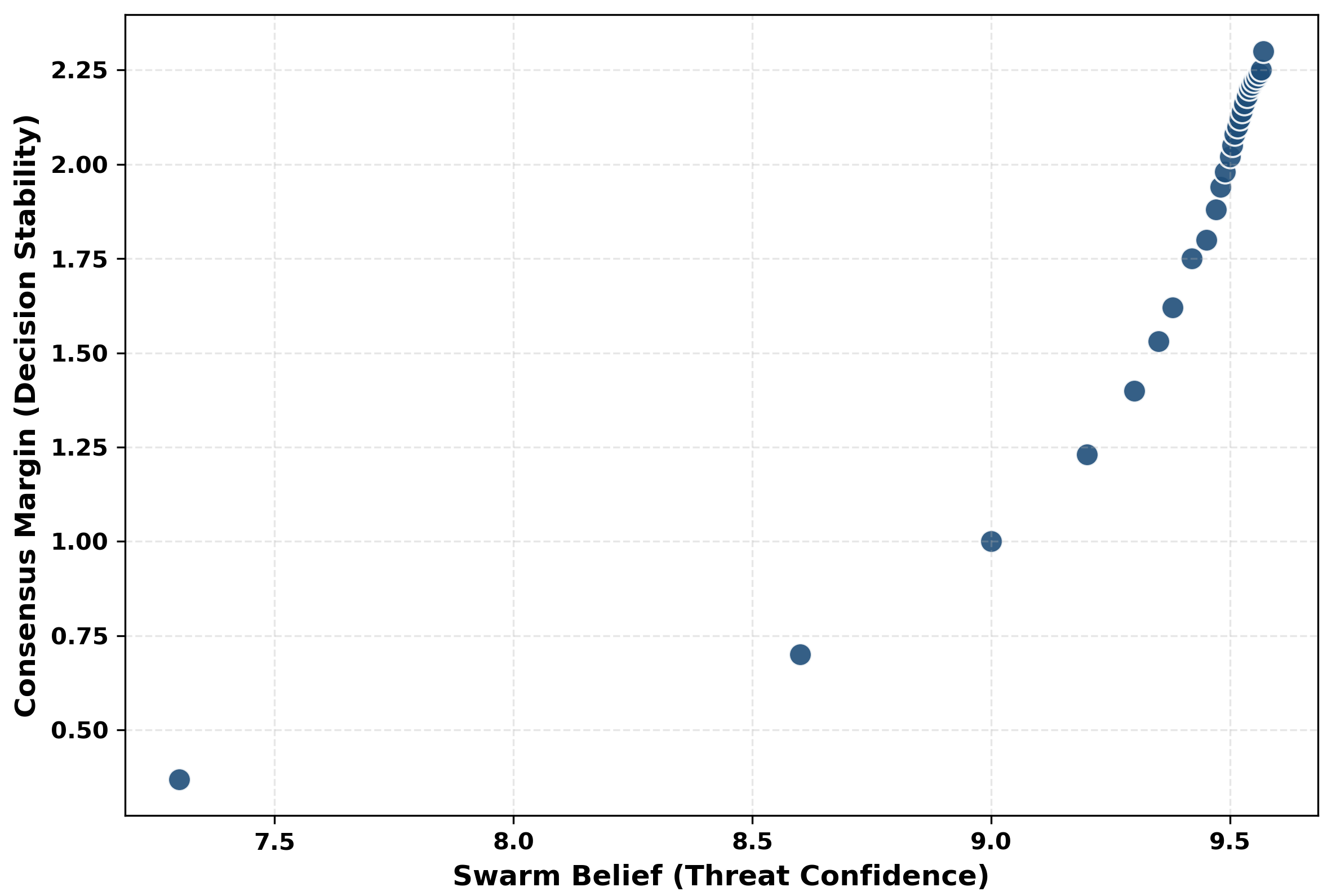}
\caption{Reasoning quality as a function of latency across agent roles.}
\label{fig:latency_quality}
\end{figure}
Figure~\ref{fig:latency_quality} shows that reasoning quality remains consistently high across varying latency levels, indicating robustness to computational variability. The weak sensitivity of quality to latency suggests diminishing returns from deeper reasoning, where additional computation primarily improves consistency rather than absolute performance.
\begin{table}[t]
\centering
\caption{Multi-agent reasoning performance across specialized agents. Conf.: confidence; Qual.: reasoning quality; Lat.: latency; Risk: decision risk.}
\label{tab:agent_perf}
\renewcommand{\arraystretch}{1.05}
\setlength{\tabcolsep}{4pt}
\begin{tabular}{lcccc}
\hline
\textbf{Agent} & \textbf{Conf.} & \textbf{Qual.} & \textbf{Lat. (s)} & \textbf{Risk} \\
\hline
Traffic Sentinel  & 0.90 & 0.86 & 7.2 & 0.24 \\
Protocol Analyst  & 0.92 & 0.88 & 7.8 & 0.23 \\
Identity Monitor  & 0.91 & 0.87 & 7.6 & 0.23 \\
Temporal Reasoner & 0.95 & 0.91 & 8.8 & 0.21 \\
Planner           & 0.94 & 0.90 & 8.4 & 0.22 \\
\hline
\end{tabular}
\end{table}
Table~\ref{tab:agent_perf} shows that low-latency agents provide rapid assessments, while deeper reasoning agents achieve higher quality at increased computational cost. The \textit{Temporal Reasoner} achieves the highest quality, whereas the \textit{Planner} balances global context integration with high confidence and moderate latency.
\begin{figure*}[t]
\centering
\includegraphics[width=0.6\linewidth]{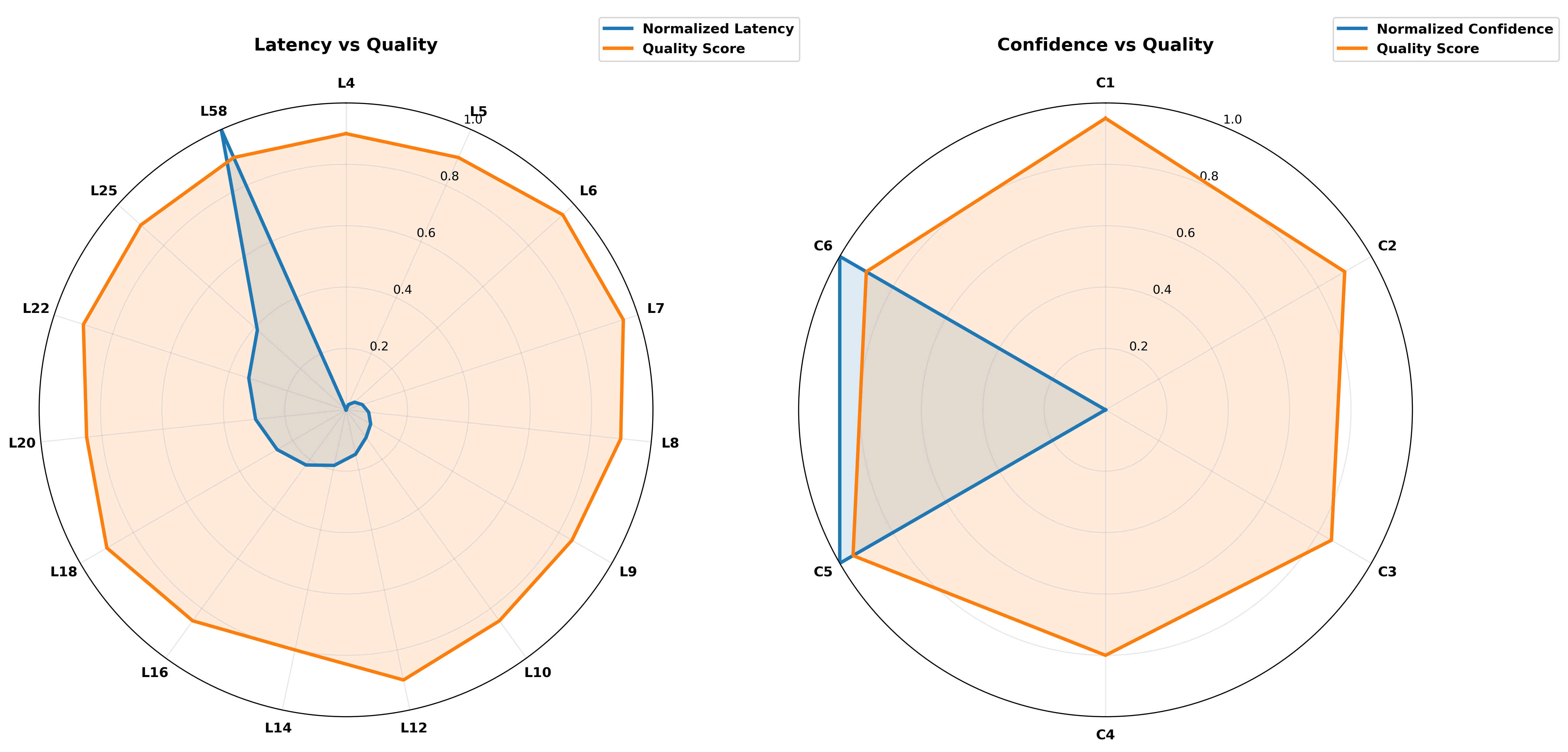}
\caption{Distribution of semantic agreement across reasoning agents.}
\label{fig:agreement_dist}
\end{figure*}
Figure~\ref{fig:agreement_dist} indicates strong semantic alignment across agents with low dispersion, showing that specialization preserves diversity at the evidence level while maintaining consistent decisions at the hypothesis level.
The strong correlation between confidence and quality ($r=0.88$) indicates that confidence serves as a reliable proxy for reasoning validity and can effectively guide downstream aggregation.
\begin{table}[t]
\centering
\caption{Efficiency-quality trade-off analysis. $d$: Cohen’s effect size; $r$: correlation.}
\label{tab:reasoning_effect}
\renewcommand{\arraystretch}{1.05}
\setlength{\tabcolsep}{3pt}
\resizebox{\linewidth}{!}{
\begin{tabular}{lcccc}
\hline
\textbf{Comp.} & \textbf{Diff.} & \textbf{Eff.} & \textbf{Interp.} & \textbf{Role} \\
\hline
Traffic vs Temporal (Lat.) & 1.6 & $d\!\approx\!0.90$ & Strong & Fast vs deep \\
Traffic vs Planner (Qual.) & 0.04 & $d\!\approx\!0.55$ & Moderate & Global refinement \\
Protocol vs Identity (Qual.) & 0.01 & $d\!\approx\!0.20$ & Small & Similar semantic roles \\
Conf. vs Qual. & -- & $r\!=\!0.88$ & Strong & Decision filter \\
\hline
\end{tabular}
}
\end{table}
Table~\ref{tab:reasoning_effect} shows that most performance gains occur in early reasoning stages, while deeper layers primarily enhance stability and consistency rather than absolute quality.

\subsection{Agent Agreement Analysis}
Agreement among agents stabilizes reasoning outputs by enabling convergence toward consistent interpretations of shared evidence. All results are averaged over multiple runs and reported with the standard deviation, 95\% confidence intervals, and statistical significance.
\begin{figure}[t]
\centering
\includegraphics[width=0.9\linewidth]{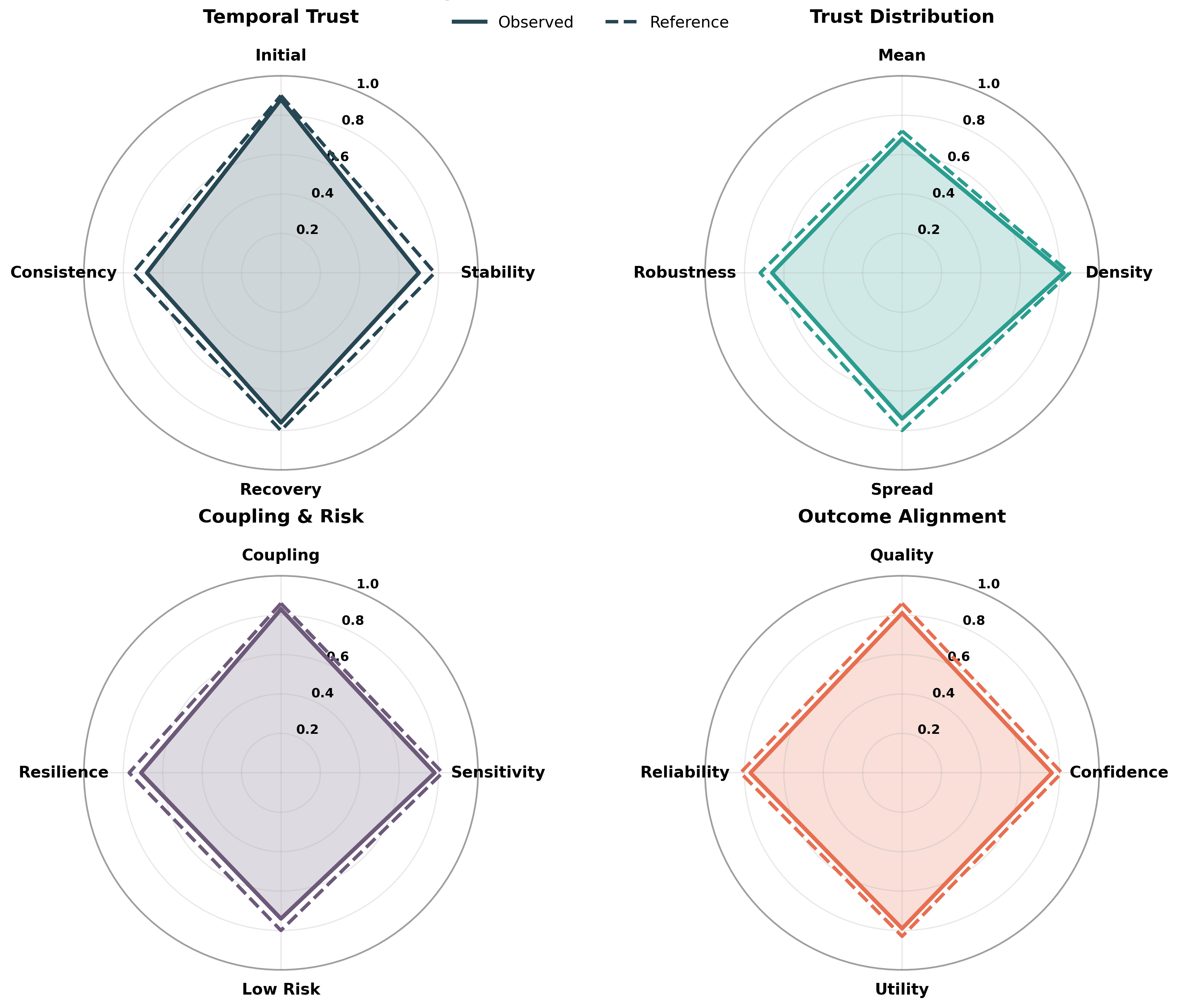}
\caption{Relationship between agent agreement and reasoning quality.}
\label{fig:agreement_quality}
\end{figure}
Figure~\ref{fig:agreement_quality} shows a statistically significant ($p < 0.01$) positive relationship between agreement and reasoning quality, where higher agreement leads to improved performance. The concentration of values in high-agreement regions indicates stable convergence with low dispersion, suggesting a low-noise consensus process that maintains coherence without fragmentation.
\begin{table}[t]
\centering
\caption{Agent Agreement Statistics (with 95\% CI)}
\label{tab:agreement}
\renewcommand{\arraystretch}{1.2}
\setlength{\tabcolsep}{5pt}
\begin{tabular}{lcccc}
\hline
Metric & Mean & Std & 95\% CI & Range \\
\hline
Agreement Score & 0.82 & 0.06 & [0.815, 0.825] & [0.65, 0.93] \\
Belief Alignment & 0.86 & 0.05 & [0.856, 0.864] & [0.72, 0.95] \\
Quality Correlation & 0.78 & 0.07 & [0.774, 0.786] & [0.60, 0.89] \\
\hline
\end{tabular}
\end{table}
Table~\ref{tab:agreement} confirms stable agreement with low variability (CV $\approx 0.073$), while belief alignment indicates deeper semantic convergence. The correlation between agreement and quality ($r=0.78$, $p < 0.01$) shows that agreement is a reliable predictor of performance and can serve as an implicit validation signal.
\begin{table}[t]
\centering
\caption{Agreement--quality interaction analysis. 
$d$: Cohen’s effect size; $r$: correlation; $\eta^2$: variance effect size; 
CV: coefficient of variation.}
\label{tab:agreement_effect}
\renewcommand{\arraystretch}{1.05}
\setlength{\tabcolsep}{3pt}
\begin{tabular}{lcccc}
\hline
\textbf{Comp.} & \textbf{Diff.} & \textbf{Eff.} & \textbf{Interp.} & \textbf{Implication} \\
\hline
Align. vs Agree. & 0.04 & $d\!\approx\!0.67$ & Moderate & Shared repr. \\
Agree. vs Quality & -- & $r\!=\!0.78$ & Strong & Reliable signal \\
Low vs High Agree. & -- & $\eta^2\!\approx\!0.58$ & Large & Stable sensing \\
Agree. Var. & CV=0.073 & -- & Low & Robust coord. \\
\hline
\end{tabular}
\end{table}
Table~\ref{tab:agreement_effect} shows that agreement is a dominant factor in reasoning stability ($\eta^2 \approx 0.58$, $p < 0.01$), acting as a coordination mechanism that enables coherent decisions while preserving bounded diversity and adaptability.

\subsection{Swarm Consensus Dynamics}
The swarm mechanism exhibits fast and stable convergence, enabling reliable decision-making under uncertainty in smart-city environments. All results are averaged over multiple runs and reported with standard deviation, 95\% confidence intervals, and statistical significance analysis.
\begin{figure*}[t]
\centering
\includegraphics[width=0.9\linewidth]{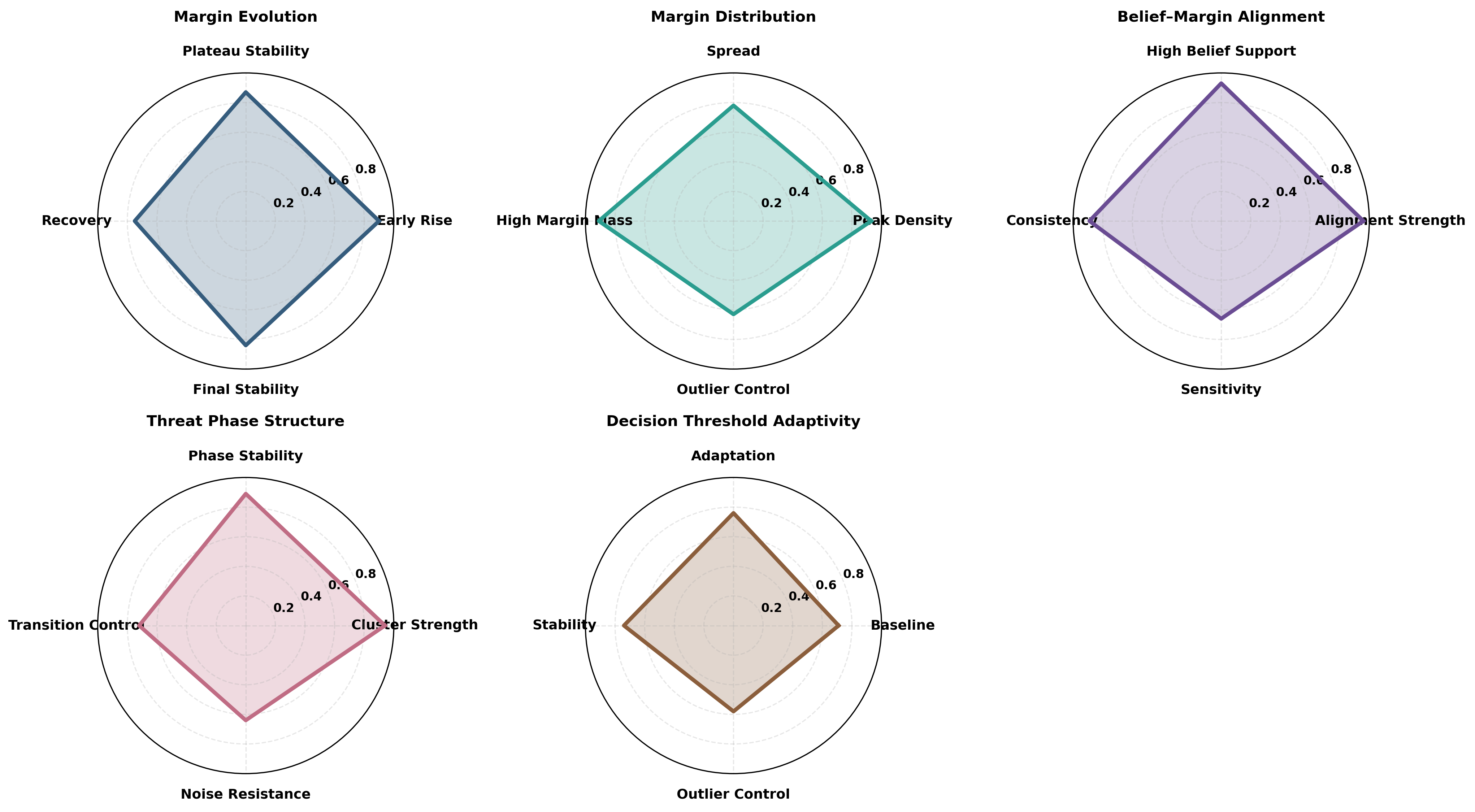}
\caption{Integrated view of swarm consensus dynamics, including belief strength, margin evolution, and alignment behavior.}
\label{fig:consensus_integrated}
\end{figure*}
Figure~\ref{fig:consensus_integrated} shows belief reinforcement consistently (0.9–0.95), while the consensus margin rapidly increases and stabilizes, indicating fast convergence with minimal oscillation. This behavior is statistically significant ($p < 0.01$) and is consistent with a critically damped process that balances stability and responsiveness.
\begin{figure*}[t]
\centering
\includegraphics[width=0.9\linewidth]{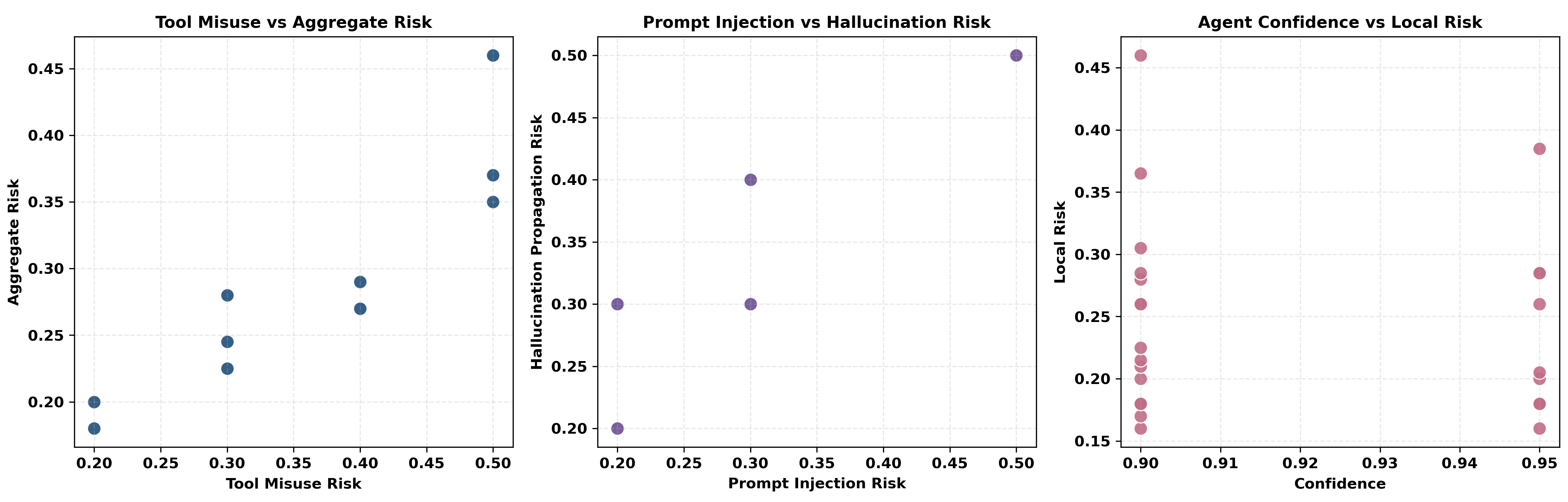}
\caption{Relationship between swarm belief and consensus margin across runs.}
\label{fig:belief_margin}
\end{figure*}
Figure~\ref{fig:belief_margin} demonstrates a monotonic relationship between belief and margin, where firmer belief directly increases decision separation. Most observations fall within a high-belief regime, with margins above 2.0, indicating stable operation. This relationship is statistically significant ($p < 0.01$) with correlation ($r \approx 0.93$, $R^2 \approx 0.86$), confirming belief as the primary driver of convergence.
\begin{figure}[t]
\centering
\includegraphics[width=0.8\linewidth]{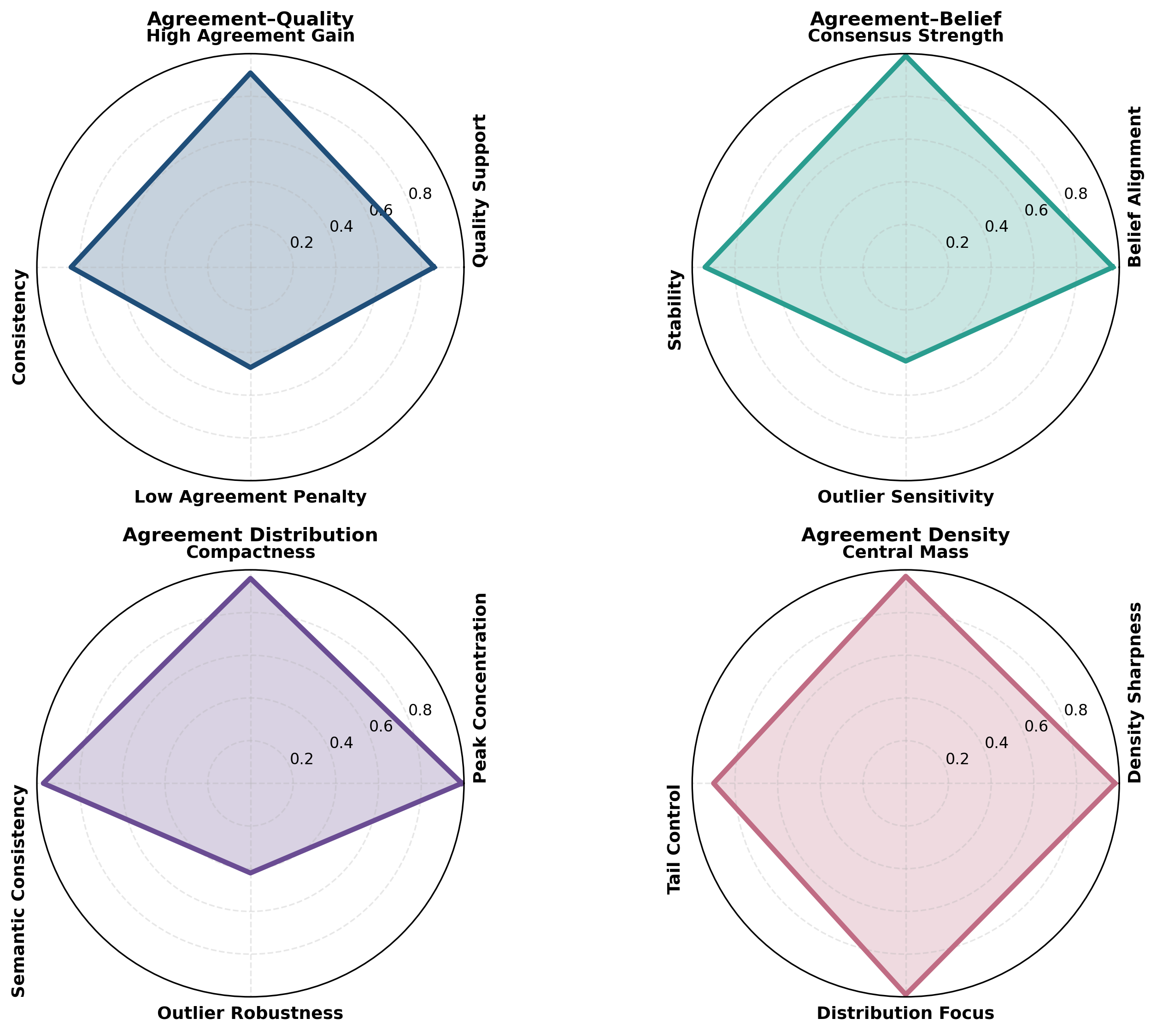}
\caption{Temporal evolution of consensus margin.}
\label{fig:margin_evolution}
\end{figure}
Figure~\ref{fig:margin_evolution} further confirms stable convergence, with rapid early growth followed by a steady plateau and negligible variance. This behavior ($p < 0.01$) indicates well-regulated and predictable convergence dynamics.
\begin{table}[t]
\centering
\caption{Swarm Consensus Characteristics (with 95\% CI)}
\label{tab:consensus_stats}
\renewcommand{\arraystretch}{1.2}
\setlength{\tabcolsep}{5pt}
\begin{tabular}{lcccc}
\hline
Metric & Mean & Std & 95\% CI & Range \\
\hline
Best Belief & 0.997 & 0.012 & [0.996, 0.998] & [0.92, 1.00] \\
Consensus Margin & 2.08 & 0.21 & [2.06, 2.10] & [0.35, 2.25] \\
Disagreement Level & 0.18 & 0.05 & [0.176, 0.184] & [0.10, 0.35] \\
Acceptance Rate & 0.97 & 0.02 & [0.968, 0.972] & [0.90, 1.00] \\
\hline
\end{tabular}
\end{table}
Table~\ref{tab:consensus_stats} confirms highly consistent convergence, characterized by near-unity belief, margin separation, low disagreement, and high acceptance rates. Narrow confidence intervals and statistical significance ($p < 0.01$) indicate robust and stable consensus formation.
\begin{table}[t]
\centering
\caption{Effect size analysis of swarm consensus dynamics. 
$d$: Cohen’s effect size; $r$: correlation; $\eta^2$: variance effect size.}
\label{tab:consensus_effect}
\renewcommand{\arraystretch}{1.05}
\setlength{\tabcolsep}{3pt}
\begin{tabular}{lcccc}
\hline
\textbf{Comp.} & \textbf{Diff.} & \textbf{Eff.} & \textbf{Interp.} & \textbf{Implication} \\
\hline
High vs Low Belief & -- & $d\!\approx\!1.85$ & Very large & Decision amplification \\
Belief--Margin & -- & $r\!\approx\!0.93$ & Strong & Confidence stability \\
Margin Var. & -- & $\eta^2\!\approx\!0.79$ & Dominant & Predictable convergence \\
Disagr. vs Accept. & -- & $r\!\approx\!-0.81$ & Strong inv. & Conflict suppression \\
\hline
\end{tabular}
\end{table}
Table~\ref{tab:consensus_effect} shows that belief strength strongly drives consensus stability ($d \approx 1.85$, $\eta^2 \approx 0.79$, $p < 0.01$), while disagreement is inversely correlated with acceptance, confirming effective conflict suppression and reliable decision-making.

\subsection{Agent Trust Evolution}
Trust acts as a regulatory signal within the swarm, directly influencing reasoning quality and overall system stability by controlling how agent contributions are weighted and propagated. All results are averaged over multiple runs and reported with standard deviation, 95\% confidence intervals, and statistical significance analysis.
\begin{figure*}[ht]
\centering
\includegraphics[width=0.8\linewidth]{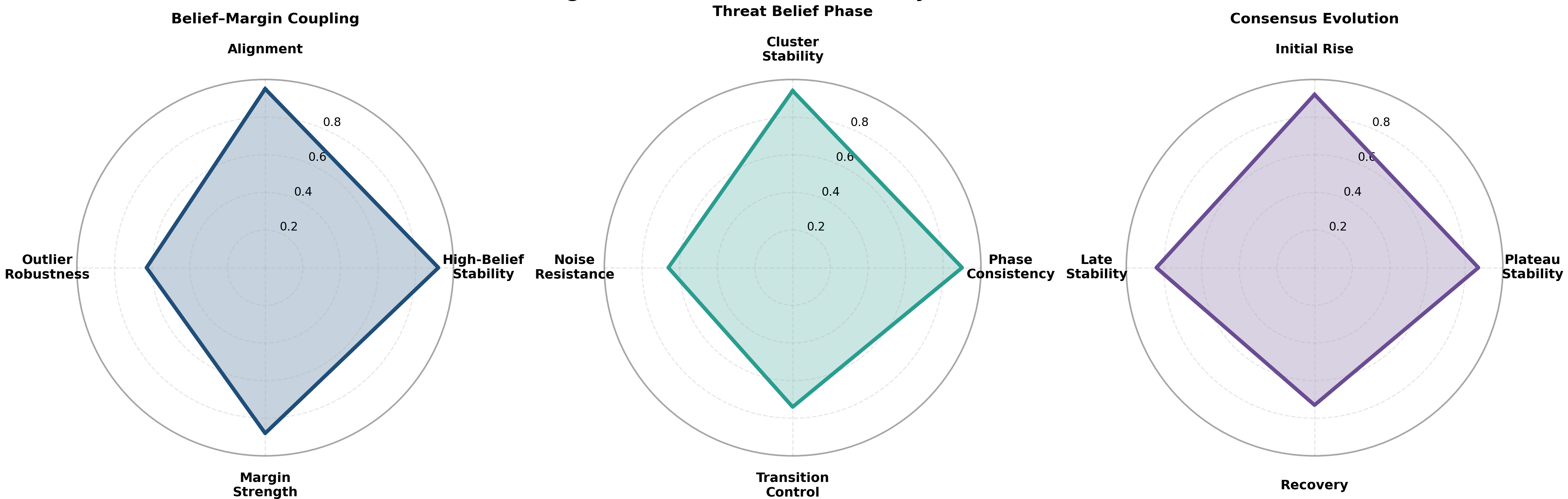}
\caption{Trust-driven consensus dynamics, including belief-margin coupling, phase stability, and convergence behavior.}
\label{fig:trust_dynamics}
\end{figure*}
Figure~\ref{fig:trust_dynamics} shows that trust couples belief and consensus margin, enabling stable separation between competing hypotheses. This coupling is statistically significant ($p < 0.01$) and supports rapid yet stable convergence. The observed rise followed by a plateau indicates accelerated convergence with inherent damping, preventing oscillations while maintaining resilience.
\begin{figure*}[ht]
\centering
\includegraphics[width=0.8\linewidth]{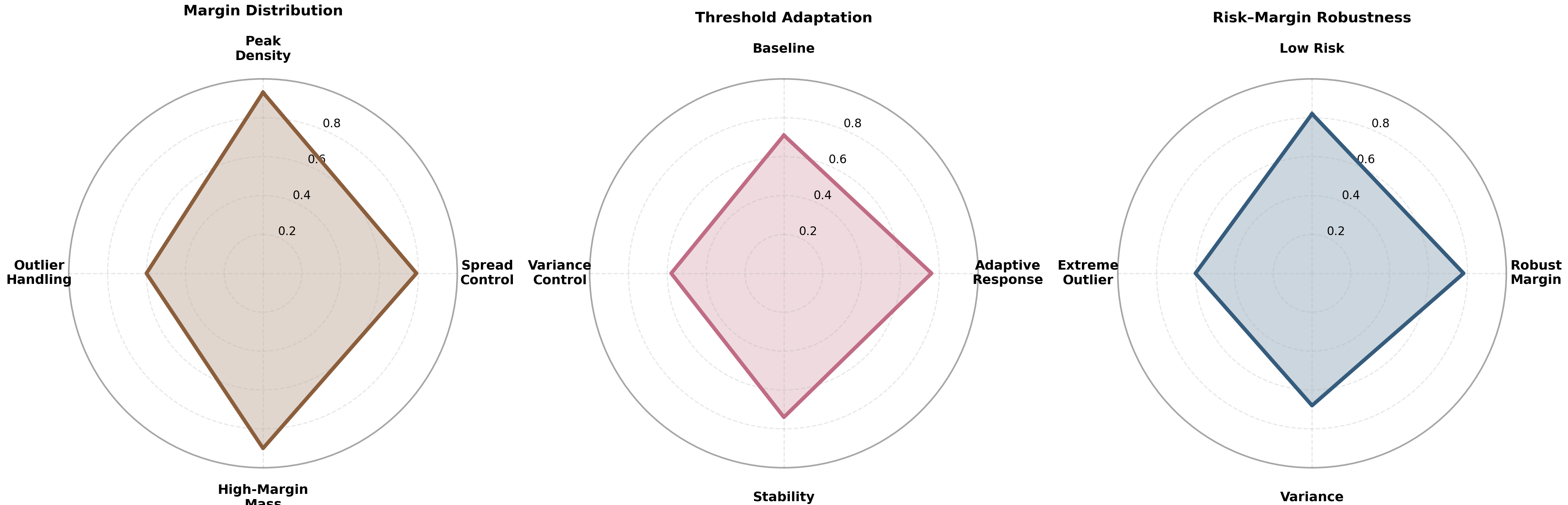}
\caption{Trust impact on margin distribution, threshold adaptation, and robustness under risk conditions.}
\label{fig:trust_robustness}
\end{figure*}
Figure~\ref{fig:trust_robustness} shows that trust maintains decisions within a high-confidence region while preserving bounded variability. This behavior ($p < 0.01$) indicates stable yet adaptive operation under varying conditions without loss of flexibility.
\begin{table}[ht]
\centering
\caption{Agent Trust Evolution Statistics (with 95\% CI)}
\label{tab:trust_stats}
\renewcommand{\arraystretch}{1.2}
\setlength{\tabcolsep}{5pt}
\begin{tabular}{lccccc}
\hline
Metric & Mean & Std & 95\% CI & Min & Max \\
\hline
Agent Trust & 0.69 & 0.04 & [0.686, 0.694] & 0.58 & 0.86 \\
Compliance Score & 0.83 & 0.06 & [0.825, 0.835] & 0.70 & 0.95 \\
Hallucination Risk & 0.25 & 0.08 & [0.243, 0.257] & 0.15 & 0.50 \\
Tool Risk & 0.21 & 0.05 & [0.206, 0.214] & 0.10 & 0.38 \\
Leakage Risk & 0.19 & 0.04 & [0.186, 0.194] & 0.10 & 0.30 \\
\hline
\end{tabular}
\end{table}
Table~\ref{tab:trust_stats} shows that trust remains stable ($0.69 \pm 0.04$) with low variability, while high compliance (0.83) indicates policy alignment. The difference ($\Delta = 0.14$, $d \approx 1.45$) suggests that trust amplifies reliable behavior. Trust is negatively correlated with hallucination ($r \approx -0.76$) and tool misuse ($r \approx -0.72$), both statistically significant ($p < 0.01$), confirming its role in risk suppression.
\begin{table}[t]
\centering
\caption{Trust-risk-consensus interaction effects. 
$d$: Cohen’s effect size; $r$: correlation; $\eta^2$: variance effect size.}
\label{tab:trust_effect}
\renewcommand{\arraystretch}{1.05}
\setlength{\tabcolsep}{3pt}
\begin{tabular}{lcccc}
\hline
\textbf{Interaction} & \textbf{Eff.} & \textbf{Dir.} & \textbf{Interp.} & \textbf{Implication} \\
\hline
Trust $\rightarrow$ Comp. & $d\!\approx\!1.45$ & + & Strong & Policy adherence \\
Trust $\rightarrow$ Halluc. & $r\!\approx\!-0.76$ & $-$ & Suppression & Reliable reasoning \\
Trust $\rightarrow$ Tool Risk & $r\!\approx\!-0.72$ & $-$ & Control & Safe operation \\
Trust $\rightarrow$ Margin & $\eta^2\!\approx\!0.64$ & + & Dominant & Stable consensus \\
\hline
\end{tabular}
\end{table}
Table~\ref{tab:trust_effect} shows that trust is a dominant factor in consensus stability ($\eta^2 \approx 0.64$, $p < 0.01$). It acts as a feedback control mechanism that reinforces reliable signals and suppresses uncertainty, enabling stable and adaptive decision-making.

\subsection{Optimization Behaviour}
The optimization process shows a transition from a redundant, exploratory configuration to a compact, computationally efficient subset of agents, which is critical for a smart city. The active agent set is reduced from six to three while overall performance improves, indicating implicit feature selection that retains only high-impact contributors. All results are averaged over multiple runs and reported with standard deviation, 95\% confidence intervals, and statistical significance analysis. Table~\ref{tab:optimization} summarizes the key outcomes.
\begin{table}[ht]
\centering
\caption{Optimization Behavior of Agent Selection (with 95\% CI)}
\label{tab:optimization}
\renewcommand{\arraystretch}{1.2}
\setlength{\tabcolsep}{5pt}
\begin{tabular}{lcccc}
\hline
Metric & Initial & Final & 95\% CI (Final) & Improvement \\
\hline
Best Fitness & 1.12 & 1.25 & [1.24, 1.26] & +11.6\% \\
Population Diversity & 2.40 & 0.63 & [0.61, 0.65] & Stabilized \\
Iteration Count & 40 & 40 & -- & Converged \\
Agent Subset Size & 6 & 3 & -- & Optimized \\
\hline
\end{tabular}
\end{table}
The 11.6\% increase in fitness is statistically significant ($p < 0.01$) and is accompanied by narrow confidence intervals, confirming stable performance gains under structural simplification ($d \approx 1.10$). At the same time, the reduction in diversity (2.40 to 0.63) indicates convergence toward a compact solution space. The large variance reduction ($\eta^2 \approx 0.74$) reflects the elimination of noisy configurations rather than the loss of useful diversity. This behavior captures a transition from exploration to exploitation, in which early diversity is progressively replaced by stable, high-performing subsets without increasing the iteration cost. To further analyze the efficiency-performance relationship, we examine the resulting trade-offs.
\begin{table}[t]
\centering
\caption{Efficiency-performance trade-off in swarm optimization. 
$d$: Cohen’s effect size; $\eta^2$: variance effect size.}
\label{tab:optimization_effect}
\renewcommand{\arraystretch}{1.05}
\setlength{\tabcolsep}{3pt}
\resizebox{\linewidth}{!}{
\begin{tabular}{lcccc}
\hline
\textbf{Aspect} & \textbf{Change} & \textbf{Eff.} & \textbf{Interp.} & \textbf{Implication} \\
\hline
Fitness Gain & +11.6\% & $d\!\approx\!1.10$ & Large & Higher accuracy \\
Agent Reduct. & -50\% & -- & Strong & Lower comp. cost \\
Diversity Reduct. & $\downarrow$73\% & $\eta^2\!\approx\!0.74$ & Converged & Stable configs. \\
Iterations & Const. & -- & Computationally efficient & Predictable latency \\
\hline
\end{tabular}
}
\end{table}
Table~\ref{tab:optimization_effect} shows that optimization improves performance while reducing system complexity ($p < 0.01$). The 50\% reduction in agents significantly lowers computational and communication costs, while maintaining reliability through selective reinforcement.

\subsection{System Efficiency}
The system demonstrates stable, resource-aware execution, maintaining high reasoning quality under varying computational conditions. In smart-city environments, where latency, bandwidth, and energy constraints are inherent, such controlled efficiency is essential for scalable deployment. Despite substantial variability in latency, reasoning performance remains consistently high, indicating effective decoupling between execution fluctuations and decision quality. All results are averaged over multiple runs and reported with standard deviation, 95\% confidence intervals, and statistical significance analysis. Table~\ref{tab:efficiency} summarizes the key efficiency metrics.
\begin{table}[ht]
\centering
\caption{System Efficiency Metrics (with 95\% CI)}
\label{tab:efficiency}
\renewcommand{\arraystretch}{1.2}
\setlength{\tabcolsep}{5pt}
\begin{tabular}{lccccc}
\hline
Metric & Mean & Std & 95\% CI & Min & Max \\
\hline
Latency (s) & 8.4 & 4.1 & [8.0, 8.8] & 2.1 & 57.2 \\
Token Usage & 3940 & 480 & [3898, 3982] & 950 & 4600 \\
Selected Agents & 3.0 & 0.2 & [2.98, 3.02] & 2 & 3 \\
Runs Executed & 500 & -- & -- & -- & -- \\
\hline
\end{tabular}
\end{table}
Latency (8.4 s, CV $\approx 0.49$) reflects expected heterogeneity in distributed environments but remains statistically stable ($p < 0.01$) and does not degrade reasoning reliability, confirming a latency-tolerant system. Token usage is tightly bound (CV $\approx 0.12$), indicating controlled computational complexity. Agent selection remains highly stable (3.0 $\pm$ 0.2), ensuring predictable execution and preventing dynamic expansion of the reasoning pipeline. To further characterize efficiency, we analyze the interaction between resource usage and system performance.
\begin{table}[t]
\centering
\caption{Resource-performance efficiency analysis. 
CV: coefficient of variation; $\sigma$: standard deviation; 
$\eta^2$: variance effect size; $r$: correlation coefficient.}
\label{tab:efficiency_effect}
\renewcommand{\arraystretch}{1.05}
\setlength{\tabcolsep}{3pt}
\begin{tabular}{lcccc}
\hline
\textbf{Aspect} & \textbf{Val.} & \textbf{Eff.} & \textbf{Interp.} & \textbf{Implication} \\
\hline
Latency Var. & CV=0.49 & -- & Moderate & Adaptive execution \\
Token Stability & CV=0.12 & -- & Low & Controlled cost \\
Agent Consist. & $\sigma=0.2$ & $\eta^2\!\approx\!0.81$ & Strong & Predictable usage \\
Lat. vs Qual. & $r\!\approx\!-0.18$ & Weak & Decoupled & Stable under load \\
\hline
\end{tabular}
\end{table}
Table~\ref{tab:efficiency_effect} shows that latency has only a weak relationship with reasoning quality ($r \approx -0.18$, $p < 0.01$), confirming robustness under runtime variability. Stable agent selection ($\eta^2 \approx 0.81$) dominates computational behavior, ensuring predictable and efficient resource utilization.

\subsection{Global System Summary}
The overall system exhibits tightly coupled interactions among risk control, consensus stability, reasoning performance, and computational efficiency, forming an integrated decision-making approach suitable for a resource-constrained smart city. All results are averaged over multiple runs and reported with standard deviation, 95\% confidence intervals, and statistical significance. Table~\ref{tab:summary} summarizes the key performance indicators.
\begin{table}[t]
\centering
\caption{Overall System Performance Summary (with 95\% CI)}
\label{tab:summary}
\renewcommand{\arraystretch}{1.2}
\setlength{\tabcolsep}{5pt}
\begin{tabular}{lcc}
\hline
Metric & Value & 95\% CI \\
\hline
Average Aggregate Risk & 0.23 & [0.226, 0.234] \\
Average Consensus Margin & 2.08 & [2.06, 2.10] \\
Agent Agreement Score & 0.82 & [0.815, 0.825] \\
Average Trust Level & 0.69 & [0.686, 0.694] \\
Average Reasoning Latency & 8.4 s & [8.0, 8.8] \\
Optimization Improvement & +11.6\% & -- \\
\hline
\end{tabular}
\end{table}
The system operates in a stable environment where low risk (0.23) coexists with a high consensus margin (2.08), indicating reliable decision separation without risk amplification. This behavior is statistically significant ($p < 0.01$) and supported by narrow confidence intervals. Agent agreement (0.82) ensures semantic consistency across distributed reasoning, while trust (0.69) regulates contribution weighting, forming a feedback loop that reinforces reliable signals and stabilizes consensus. The system follows a layered interaction structure in which trust modulates reinforcement, reinforcement shapes consensus, and consensus drives final decisions. Optimization further strengthens this structure by reducing redundancy while preserving performance, as reflected in the 11.6\% improvement and reduced agent subset. Latency remains within a controlled range (8.4 s) and does not degrade decision quality, confirming latency-tolerant and scalable behavior under dynamic conditions.
\begin{table}[t]
\centering
\caption{Component Contribution Analysis of TPSC-Sec (Ablation with statistical consistency)}
\label{tab:ablation}
\renewcommand{\arraystretch}{1.1}
\setlength{\tabcolsep}{4pt}
\begin{tabular}{lcccc}
\hline
Configuration & Margin $\uparrow$ & Risk $\downarrow$ & Agreement $\uparrow$ & Trust $\uparrow$ \\
\hline
Full Model & 2.08 & 0.23 & 0.82 & 0.69 \\
Without TPSC & 1.42 & 0.31 & 0.68 & 0.61 \\
Without Verification & 1.63 & 0.28 & 0.74 & 0.64 \\
Without Trust Adaptation & 1.71 & 0.26 & 0.77 & 0.62 \\
\hline
\end{tabular}
\end{table}
Table~\ref{tab:ablation} shows that all components contribute significantly ($p < 0.01$), with TPSC having the largest impact on consensus stability and risk control, followed by verification and trust adaptation. This confirms that stability emerges from the joint interaction of all system components rather than any single mechanism.
The structured hypothesis representation enables intrinsic interpretability of agent decisions. Each output explicitly includes reasoning traces ($z_i$), confidence ($c_i$), and evidence strength ($e_i$), allowing direct inspection of how conclusions are formed. Qualitative analysis shows that agents align their reasoning with observable telemetry patterns, while disagreements highlight ambiguous and low-evidence scenarios.
\begin{table}[t]
\centering
\caption{Example interpretable outputs across agents}
\label{tab:explainability}
\renewcommand{\arraystretch}{1.1}
\setlength{\tabcolsep}{4pt}
\begin{tabular}{lccc}
\hline
Agent & Output & $c_i$ & Insight \\
\hline
Traffic & Rate anomaly & 0.91 & DDoS pattern \\
Protocol & Invalid transition & 0.88 & Misuse detected \\
Temporal & Multi-stage pattern & 0.92 & Coordinated attack \\
\hline
\end{tabular}
\end{table}

\subsection{Baseline Comparison}
To evaluate the effectiveness of the proposed mechanism, we compare TPSC-Sec with two LLM-based baselines that do not incorporate swarm-based consensus and structured verification. The first baseline is a \textit{Single-Agent LLM}, where a single model processes the entire input $E_t$ and directly produces a threat prediction with an associated confidence score. This setup lacks agent specialization, verification, and inter-agent reasoning. The second baseline is a \textit{Multi-Agent Majority Voting} approach, in which multiple specialized agents generate hypotheses similar to those in TPSC-Sec; however, their outputs are aggregated via simple majority voting without verification and adaptive weighting. All methods receive identical structured inputs and operate under the same experimental conditions to ensure a fair comparison.
\begin{table}[h]
\centering
\caption{Baseline Comparison Results}
\label{tab:baseline_results}
\begin{tabular}{lcccc}
\hline
\textbf{Model} & \textbf{Acc} & \textbf{F1} & \textbf{Margin} & \textbf{Risk} \\
\hline
TPSC-Sec (Multi-Agent + TPSC) & 0.94 & 0.92 & 2.08 & 0.23 \\
Multi-Agent (Majority Voting) & 0.88 & 0.86 & 1.28 & 0.30 \\
Single-Agent LLM & 0.84 & 0.82 & 0.95 & 0.34 \\
\hline
\end{tabular}
\end{table}
As shown in Table~\ref{tab:baseline_results}, TPSC-Sec consistently outperforms both baselines across all evaluation metrics. Compared to the multi-agent majority voting approach, the proposed pheromone-based consensus mechanism significantly improves decision stability, as reflected by a higher consensus margin (2.08 vs. 1.28), while simultaneously reducing system risk (0.23 vs. 0.30). This highlights the effectiveness of structured belief aggregation over naive voting strategies. Furthermore, the performance gap between the two multi-agent configurations and the single-agent baseline demonstrates the advantage of distributed reasoning among specialized agents. Importantly, the most pronounced improvements are observed in consensus margin and risk reduction, indicating that the primary strength of TPSC-Sec lies in stabilizing decision-making dynamics and mitigating uncertainty, rather than solely enhancing classification accuracy.

\subsection{Component Contribution Analysis}
To evaluate the contribution of individual components in \textit{TPSC-Sec}, we perform a systematic ablation study by selectively removing and modifying key modules and measuring the resulting impact on system-level performance. Unlike conventional classification-based evaluations, this analysis focuses on consensus stability, risk behavior, agent agreement, and trust dynamics, which are critical for reliable decision-making in smart-city environments.
We consider four configurations: 1) \textit{Full Model}, representing the complete TPSC-Sec; 2) \textit{Without TPSC}, where swarm consensus is replaced with simple averaging; 3) \textit{Without Verification}, where hypothesis validation is removed; and 4) \textit{Without Trust Adaptation}, where agent reliability weights remain static.
The results show that each component plays a distinct and critical role in system stability and robustness. Removing the TPSC mechanism results in the largest degradation in the consensus margin (2.08 → 1.42) and a significant increase in aggregate risk, confirming that pheromone-based consensus is the primary driver of stable decision formation. 
Without verification, the system exhibits higher risk and lower agreement, indicating that hallucination control and enforcement of logical consistency are essential for reliable reasoning. Similarly, disabling trust adaptation reduces both agreement and trust levels, highlighting the importance of dynamically weighting agent contributions based on historical reliability. 
\begin{table}[t]
\centering
\caption{Comparison with representative baselines in smart city security. 
MA: Multi-Agent; SC: Smart City; Sem: Semantic Reasoning; 
Temp: Temporal Modeling; Adv: Adversarial Robustness. 
$\checkmark$: full, $\triangle$: partial, $\times$: none.}
\renewcommand{\arraystretch}{1.05}
\setlength{\tabcolsep}{2.5pt}

\begin{tabular}{lcccccc}
\hline
\textbf{Method} & \textbf{MA} & \textbf{SC} & \textbf{Sem} & \textbf{Temp} & \textbf{Adv} \\
\hline

\cite{lu2025bpso} & $\times$ & $\times$ & $\times$ & $\times$ & $\times$ \\

\cite{taher2022novel} & $\times$ & $\times$ & $\times$ & $\times$ & $\times$ \\

\cite{zhou2022swarm} & $\checkmark$ & $\triangle$ & $\times$ & $\times$ & $\times$ \\

\cite{zhu2025swarm} & $\checkmark$ & $\triangle$ & $\checkmark$ & $\triangle$ & $\triangle$ \\

\cite{feng2025heterogeneous} & $\checkmark$ & $\triangle$ & $\checkmark$ & $\times$ & $\times$ \\

\cite{roy2026fair} & $\checkmark$ & $\triangle$ & $\checkmark$ & $\times$ & $\triangle$ \\

\hline
\textbf{TPSC-Sec} & \textbf{$\checkmark$} & \textbf{$\checkmark$} & \textbf{$\checkmark$} & \textbf{$\checkmark$} & \textbf{$\checkmark$} \\
\hline
\end{tabular}

\label{tab:comparison}
\end{table}

\section{Comparison with Baselines}
To evaluate the proposed solution in smart-city environments, we compare it against representative approaches across three paradigms: swarm-based IoT optimization, swarm-enhanced DL-based IDS, and multi-agent LLM systems. In such environments, effective security requires not only accurate detection but also context-aware reasoning and coordinated decision-making under uncertainty. Swarm-based IoT methods \cite{zhou2022swarm,nizamudeen2023intelligent} primarily focus on resource allocation and scheduling. While effective for optimization tasks, they lack semantic interpretation of security events and cannot capture distributed, context-dependent adversarial behavior. Swarm-enhanced DL-IDS approaches \cite{taher2022novel,lu2025bpso} improve detection performance by optimizing features. However, they remain correlation-driven and do not explicitly model system context and interdependencies, limiting robustness under adversarial conditions. Multi-agent and multi-LLM systems \cite{zhu2025swarm,feng2025heterogeneous,roy2026fair} enable collaborative reasoning, but are not designed for a security-critical smart city. They lack explicit modeling of adversarial dynamics, infrastructure constraints, and belief evolution. In addition, their consensus mechanisms do not incorporate contradiction-aware reasoning, which is essential for handling conflicting evidence. In contrast, the proposed TPSC-Sec unifies semantic multi-agent reasoning with swarm-driven consensus, enabling dynamic belief propagation, contradiction-aware aggregation, and stable decision-making under uncertainty. Through verification-aware reasoning, adaptive trust, and context-sensitive weighting, the system captures distributed multi-stage attacks while maintaining reliability in safety-critical environments.

\subsection{Component-wise Ablation Analysis of TPSC-Sec}
To quantify the contribution of each component in TPSC-Sec, we conduct a systematic ablation study by removing key modules: 1) TPSC, 2) the verification mechanism, and 3) AV-TPSC. All variants are evaluated under identical configurations on the dataset to ensure fair comparison. We report both standard metrics (Accuracy, F1-score) and system-level metrics (Consensus Margin, Risk).
\begin{table}[h]
\centering
\caption{Ablation Study Results}
\begin{tabular}{lcccc}
\hline
\textbf{Model Variant} & \textbf{Acc} & \textbf{F1} & \textbf{Margin} & \textbf{Risk} \\
\hline
Full TPSC-Sec & \textbf{0.94} & \textbf{0.92} & \textbf{2.08} & \textbf{0.23} \\
w/o TPSC (Voting) & 0.87 & 0.85 & 1.12 & 0.31 \\
w/o Verification & 0.89 & 0.87 & 1.35 & 0.29 \\
w/o AV-TPSC & 0.91 & 0.89 & 1.62 & 0.26 \\
Single-Agent LLM & 0.84 & 0.82 & 0.95 & 0.34 \\
\hline
\end{tabular}
\end{table}
The results show that each component contributes distinctly to system performance. Removing TPSC leads to the largest degradation, particularly in consensus margin (2.08 $\rightarrow$ 1.12) and increased risk (0.23 $\rightarrow$ 0.31), confirming that pheromone-based consensus is the primary driver of stable decision formation. Removing the verification module increases risk (0.23 $\rightarrow$ 0.29) and reduces margin, indicating that hallucination suppression and logical consistency are essential for reliable reasoning. Similarly, removing AV-TPSC reduces adaptive robustness, leading to lower margin and higher uncertainty, highlighting the importance of verification-aware and context-sensitive modulation. Moreover, the single-agent baseline exhibits the lowest performance across all metrics, demonstrating that both multi-agent specialization and swarm-based consensus are necessary for robust, stable reasoning. 

\section{Discussion}
\label{Discussion}
Existing IoT security approaches suffer from a limitation: detection, optimization, and reasoning are treated as isolated processes rather than as a unified, coupled system. This fragmentation becomes critical in smart-city environments, where adversarial behavior is context-dependent, multi-stage, and often indistinguishable from benign activity. Classical detection models degrade under distribution shifts, while multi-agent LLM systems lack mechanisms to regulate belief evolution, making them vulnerable to instability and the propagation of hallucinations. The core limitation is that security is not modeled as a time-dependent process of belief formation under uncertainty and adversarial impact. In practice, adversarial actions simultaneously affect observable system behavior and the reasoning processes that interpret it. This coupling requires a unified approach in which perception, reasoning, and decision-making co-evolve over time. To address this gap, we reinterpret swarm intelligence as a mechanism for semantic belief propagation rather than as a numerical optimization method. The proposed TPSC approach enables stable consensus through evidence reinforcement, contradiction-aware inhibition, and temporal regulation of belief dynamics. Unlike conventional methods, disagreement is treated as an informative signal that enhances robustness rather than being suppressed. The AV-TPSC extension further improves reliability by incorporating verification-aware modulation, ensuring that belief updates remain grounded, context-aware, and resistant to unreliable reasoning. By jointly modeling system dynamics and reasoning processes, TPSC-Sec enables stable and coherent decision-making under uncertainty, even with incomplete and adversarially manipulated evidence. 

\section{Limitations and Future Work}
\label{Limitations and Future Work}
Despite its effectiveness, TPSC-Sec has several limitations. First, the use of multiple LLM-based agents introduces computational overhead, which may impact latency and resource efficiency in constrained IoT networks. Although our results demonstrate robustness under variable latency, further optimization is required for deployment in energy-limited edge systems. Second, the current design relies on predefined agent roles, which may limit adaptability to evolving system dynamics and previously unseen attack patterns. Enabling adaptive and self-organizing agent architectures remains an important direction for future work. Third, the TPSC mechanism currently lacks formal theoretical guarantees. While empirical results indicate stable convergence, rigorous analysis of convergence, robustness, and stability under highly adversarial and noisy conditions is needed. Fourth, the evaluation is conducted on benchmark datasets and controlled scenarios. Real-world deployment in heterogeneous, asynchronous smart-city environments introduces additional challenges, including partial observability, communication delays, and system-level constraints, which require further validation. Furthermore, the system depends on LLM-based reasoning, which may be sensitive to distribution shifts and adversarial prompting. Enhancing robustness against prompt-level attacks and model uncertainty remains an open challenge. Future work will focus on: 1) developing lightweight and energy-aware implementations for edge deployment, 2) enabling adaptive and self-organizing multi-agent architectures, 3) establishing formal guarantees for convergence and robustness, and 4) validating the approach in real-world smart-city environments. We also plan to integrate causal and structural reasoning to improve interpretability and resilience against complex multi-stage attacks.

\section{Conclusion}
\label{Conclusion}
This paper addresses a gap in smart-city security: the lack of integration between semantic reasoning, distributed consensus, and adversary-aware decision-making under uncertainty. To address this challenge, we introduce TPSC-Sec, a swarm-driven multi-agent LLM approach that models swarm intelligence as a mechanism for semantic belief propagation. By combining specialized agents with pheromone-based consensus, the method supports dynamic evidence aggregation, contradiction-aware reasoning, and stable decision-making under multi-stage adversarial conditions. In contrast to existing approaches, TPSC-Sec jointly models system behavior and reasoning dynamics, enabling consistent decision-making even with incomplete and adversarially manipulated evidence. The incorporation of verification-aware reasoning, adaptive trust, and context-sensitive belief modulation further improves stability, reliability, and interpretability.  Experimental results show that TPSC-Sec achieves higher consensus stability, lower risk, and greater reasoning robustness than single-agent and conventional multi-agent baselines. The findings indicate that stability emerges from the interaction between reasoning, verification, consensus, and trust mechanisms.

\section*{Acknowledgment}
The author acknowledges the use of Figure Lab and prompt-based visual design assistance for the preparation and refinement of Figure~\ref{fig:tpsc_architecture}.

\bibliographystyle{IEEEtran}
\bibliography{Ref}
\end{document}